\documentclass[prb,twocolumn,showpacs,superscriptaddress,amsmath,amssymb,notitlepage,aps]{revtex4}
\usepackage{graphicx}
\usepackage{dcolumn}
\usepackage{amsmath,bm}
\usepackage{epstopdf}
\usepackage[mathlines]{lineno}
\usepackage{color}
\usepackage{pifont}

\usepackage[colorlinks=true,linkcolor=blue,citecolor=blue]{hyperref}
\usepackage[normalem]{ulem}
\usepackage{txfonts}
\usepackage{soul,xcolor}
\usepackage{ulem}

\usepackage{xcolor,cancel}

\def\sig{{\mbox{\boldmath{$\sigma$}}}}

\newcommand{\up}{\uparrow}
\newcommand{\down}{\downarrow}

\usepackage{soul,xcolor}
\setstcolor{red}

\begin{document}

\title{Thermoelectric performance of  nano junctions subjected to microwave driven spin-orbit coupling}

\author{Debashree  Chowdhury}
\email{debashreephys@gmail.com}
\affiliation{Centre for Nanotechnology, IIT Roorkee, Roorkee, Uttarakhand  247667, India }

\author{O. Entin-Wohlman}
%\email{orawohlman@gmail.com}
\affiliation{School of Physics and Astronomy, Tel Aviv University, Tel Aviv 6997801, Israel}

\author{A. Aharony}
\affiliation{School of Physics and Astronomy, Tel Aviv University, Tel Aviv 6997801, Israel}

\begin{abstract}

Coherent charge and heat transport through periodically driven nanodevices provide a platform for studying thermoelectric effects on the nanoscale. Here we study a junction comprising
a quantum dot connected to two fermionic terminals by two weak links. An AC electric field induces time-dependent spin-orbit interaction in the weak links. We show that this setup supports DC charge and heat currents and that thermoelectric performance can be improved, as reflected by the effect of the spin-orbit coupling on the Seebeck coefficient and the electronic thermal conductance. Our analysis is based on the nonequilibrium Keldysh Green's function formalism in the time domain
and reveals an interesting distribution of the power supply from the AC source among the various components of the device, apparently not realized before.

\end{abstract}
\date{\today}
\maketitle

%%%%%%%%%%%%%%%%%%%%%%%%%%%%%%%%%%%%	
%%%%%%%%%%%%%%%%%%%%%%%%%%%%%%%%%%%%
\section{Introduction}
\label{Intro}
%%%%%%%%%%%%%%%%%%%%%%%%%%%%%%%%%%%%	
%%%%%%%%%%%%%%%%%%%%%%%%%%%%%%%%%%%%
The interrelation between particle and energy currents is one of the main issues of quantum thermoelectrics \cite{Pekola2015, Millen2016, Benenti2017}, in particular in devices based on charge carriers. In those, the energy dependence of electronic transport is crucial for efficient thermoelectric performance. For this reason, experimental studies focus on measurements of the Seebeck (or Peltier) effect, for instance in molecular junctions \cite{Reddy2007}, or in quantum-dot heat engines \cite{Dutta2019}
(a large Seebeck coefficient enhances the thermoelectric efficiency). There are also measurements of
 the thermal conductance of a single-molecule junction \cite{Cui2019} (smaller thermal conductances favor higher figures of merit). Other experimental setups aimed at studying thermoelectric properties of single-atom heat engines realized on a calcium ion in a tapered ion trap \cite{Rosnagel2016},  of devices based on Kondo resonances in quantum dots \cite{Svilans2018, Daroca2023}, and on a quantum dot embedded into a semiconductor nanowire \cite{Josefsson2018}. A quantum heat engine was also constructed by an ensemble of nitrogen-vacancy centers in diamond \cite{Klatzow2019}.

%%%%%%%%%%%%%%%%%%%%%%%%%%%%%%%%%%%%	
%%%%%%%%%%%%%%%%%%%%%%%%%%%%%%%%%%%%

The quest for improved performance of thermoelectric electronic devices is an ongoing endeavor (see, e.g., Refs. \onlinecite{Humphery2005, Gallego2014, Vinjanampathy2016, Sanchez2019, Arrachea2023} and references therein).
In a seminal paper, Mahan and Sofo \cite{Mahan1996} proposed that high thermoelectric efficiencies of a two-terminal electronic device are obtained when the energy-dependent conductance has a sharp structure. % away from the common chemical potential of the two terminals.
A very interesting structure of the thermopower has been found in an interacting multi-level quantum dot coupled to two electronic reservoirs \cite{Beenakker1992}.
Reference \cite{Edwards1995} presented an early, ingenious way to cool a finite two-dimensional electron gas (which plays the role of the thermal bath) at low temperatures by elastic electron transitions to and from the leads. %All the energies involved are only of order $k^{}_{\rm B}T$, i.e., for $\beta\equiv (k^{}_{\rm B}T)^{-1}<1$.
%An experimental realization of some of these suggestions can be found in Ref. \cite{Prance2009}.
However, it has been found that quantum mechanics places an upper bound on both power output and efficiency at any finite power \cite{Whitney2014}.
Recall that usual heat engines reach the Carnot limit for reversible processes,  i.e., at zero entropy production, and then they produce vanishingly small outpower.
This convention was recently defined in Ref. \onlinecite{Ryo2022}, which obtained efficiencies higher than the Carnot limit
with periodically-driven chiral conductors. The surprising result hinges upon using a chiral (i.e., topological) conductor, which allows electrons moving in it to sense the effect of an AC external field depending on their direction of motion. Upon averaging over a single cycle of the AC field, no net input power is supplied to a heat engine based on topological conductors. This idea may be related to Thouless' paper \cite{Thouless1983},
which showed that a slow periodic variation of the potential landscape can yield quantized and nondissipative particle transport in unbiased junctions.

In recent studies of quantum thermodynamics the classical thermodynamic cycle is replaced by an AC time-dependent external  driving \cite{Platero2004, Zhou2015, Bruch2016, OEW2017}, in particular, periodic modulations
of the shape of quantum dots or of the potential landscape of mesoscopic junctions which turns the coherent electronic transport to be inelastic \cite{Ludovico2016, Cavina2017, Brandner2020, Bhandari2020, Saito2020, Potanina2021}.
The time-dependent variation can be assigned to the baths, i.e., to the
chemical potentials of the terminals or to their (different) temperatures \cite{Tatara2015, Lopez2023}, or to the coherent quantum system (``working substance") \cite{Ludovico2016}. In this work, we consider a quantum dot coupled to two electronic baths
by two weak links in which a spin-orbit interaction is induced by an external AC electric field. It is well-known that the
Rashba interaction  \cite{Rashba} can be tuned electrostatically \cite{Nitta1997, Sato2001, Beukman2017}. We propose that making this interaction vary periodically with time lends it added value.      Put differently,
imposing a periodically varying time-dependent spin-orbit coupling on a nanostructure thermoelectric device offers another means of enhancing its performance.

%%%%%%%%%%%%%%%%%%%%%%%%%%%%%%%%%%%%	
%%%%%%%%%%%%%%%%%%%%%%%%%%%%%%%%%%%%

%%%%%%%%%%%%%%%%%%%%%%%%%%%%%%%%%%%%	
%%%%%%%%%%%%%%%%%%%%%%%%%%%%%%%%%%%%
\begin{figure}
\includegraphics[width=0.43\textwidth]{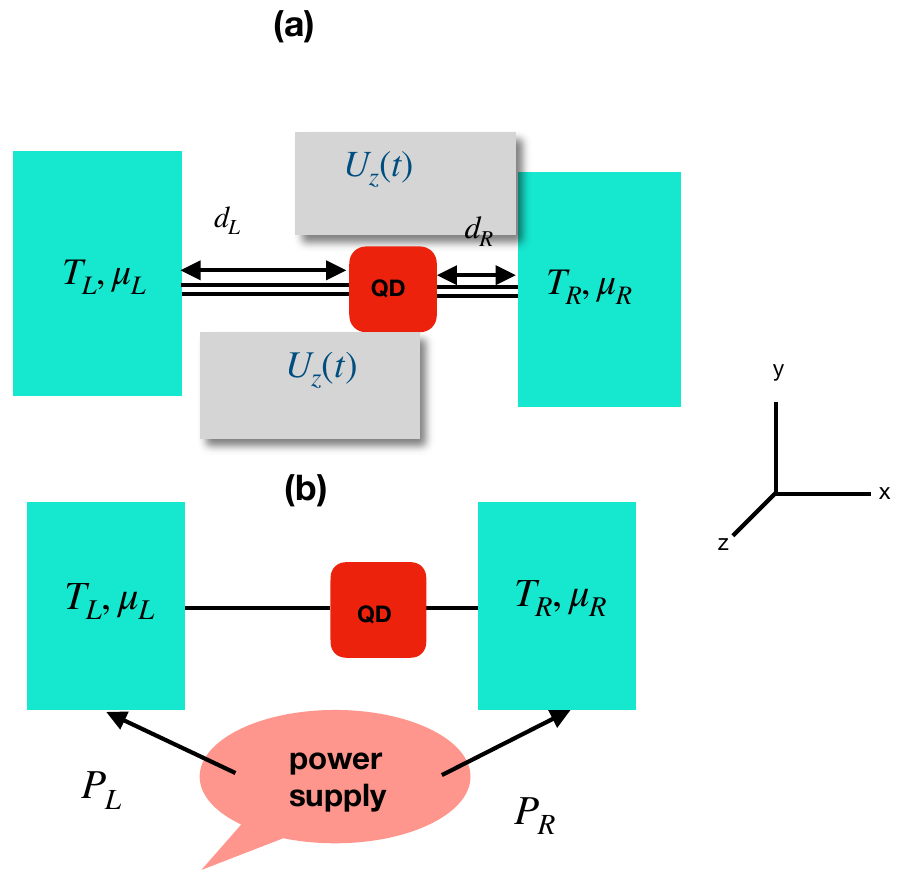}
\caption{(color online) (a) Schematic plot of the model system: a single-level (of energy $\varepsilon^{}_{d}$) quantum dot is attached by two weak links (of lengths $d^{}_{L,R}$) lying along $\hat{\bf x}$ with two electrons' baths, denoted $L$ and $R$. An electric AC field $U^{}_{z}(t)$ along $\hat{\bf z}$, whose amplitude oscillates with frequency $\Omega$, induces a Rashba spin-orbit interaction in the links. The two fermionic terminals are held at different temperatures ($T^{}_{L,R}$) and different chemical potentials ($\mu^{}_{L,R}$). (b) Upon averaging the charge and energy fluxes over a period of the AC field, the AC field supplies the power $P^{}_{L,R}$ to the left (right) terminals.}
\label{sys}
\end{figure}
%%%%%%%%%%%%%%%%%%%%%%%%%%%%%%%%%%%%	
%%%%%%%%%%%%%%%%%%%%%%%%%%%%%%%%%%%%

%%%%%%%%%%%%%%%%%%%%%%%%%%%%%%%%%%%%	
%%%%%%%%%%%%%%%%%%%%%%%%%%%%%%%%%%%%

%%%%%%%%%%%%%%%%%%%%%%%%%%%%%%%%%%%%	
%%%%%%%%%%%%%%%%%%%%%%%%%%%%%%%%%%%%

The specific setup we have in mind is depicted in Fig. \ref{sys}(a): a single-level (of energy $\varepsilon_{d}$) quantum dot, coupled to two fermionic reservoirs by spin-orbit active weak links. Had this spin-orbit coupling been static, it would have had (almost) no effect on time-independent electronic  (charge and energy) transport, because spin-orbit interaction preserves time-reversal symmetry \cite{Bardarson}:  a static spin-orbit coupling, which results in a unitary evolution of the spinor wave function, does not modify the DC transport.
However, time-reversal symmetry is destroyed by time dependence:
due to the Aharonov-Casher effect \cite{AC}, the tunneling amplitudes on the weak links attain time-dependent phase factors \cite{Meir1989, OEW2020}, which are unitary matrices (in spin space).{\color{red}\cite{comsoi}}

%%%%%%%%%%%%%%%%%%%%%%%%%%%%%%%%%%%%	
%%%%%%%%%%%%%%%%%%%%%%%%%%%%%%%%%%%%

The Aharonov-Casher phase factor dominating the weak links in Fig. \ref{sys}(a) is derived as follows. The Rashba interaction in the weak links is induced by external electric fields, which can be polarized in various ways. Here we focus on the simplest configuration of an oscillating (with frequency $\Omega$) longitudinal field,  $U_{z}(t)$, polarized along $\hat{\bf z}$.  In that case, the spin-orbit interaction appearing in the weak link is governed by the Rashba  Hamiltonian
\begin{align}
{\cal H}^{}_{\rm so}(t)=\frac{k^{}_{\rm so}}{m^{\ast}}\cos(\Omega t)\hat{\bf z}\cdot\sig\times{\bf k}\ ,
\end{align}
where the components of $\sig$ are the Pauli matrices, the spin precession wave vector $k^{}_{\rm so}$  measures the spin-orbit coupling strength (in momentum or inverse length units, using $\hbar=1$) induced by the AC field, $m^{\ast}$ is the electron's effective mass in the weak link,  and ${\bf k}=\hat{\bf x}k$ is the electron's wave-vector.
Adding the Hamiltonian of a free electron, ${\cal H}_{0}=k^{2}/(2m^{\ast})$, the propagator along the left (right) link \cite{Shahbazyan1994, Entin2019} acquires  Aharonov-Casher phase factors \cite{AC},
$\exp[i\varphi^{}_{L(R)}(t)]$,
\begin{align}
\exp[i\varphi^{}_{L}(t)]&=e^{ik^{}_{\rm so}d^{}_{L}\cos(\Omega t)\hat{\bf x}\times\hat{\bf z}\cdot\sig}=e^{-ik^{}_{\rm so}d^{}_{L}\cos(\Omega t)\sigma^{}_{y}}\ ,\nonumber\\
\exp[i\varphi^{}_{R}(t)]&=e^{ik^{}_{\rm so}d^{}_{R}\cos(\Omega t)\hat{\bf z}\times\hat{\bf x}\cdot\sig}=e^{ik^{}_{\rm so}d^{}_{R}\cos(\Omega t)\sigma^{}_{y}}\ ,
\label{acL}
\end{align}
which are unitary matrices in spin space.~ {\color{red}\cite{comsoi}}

%%%%%%%%%%%%%%%%%%%%%%%%%%%%%%%%%%%%	
%%%%%%%%%%%%%%%%%%%%%%%%%%%%%%%%%%%%

The Aharonov-Casher phase factors are incorporated into the tunneling Hamiltonian pertaining to the weak links connecting  the quantum dot with the electrons' baths
\begin{align}
{\cal H}^{}_{\rm tun}(t)&=J^{}_{L}\sum_{\bf k}c^{\dagger}_{\bf k}\exp[i\varphi^{}_{L}(t)]c^{}_{d}
\nonumber\\
&+J^{}_{R}\sum_{\bf p}c^{\dagger}_{\bf p}\exp[i\varphi^{}_{R}(t)]c^{}_{d}+{\rm H.c.}\ ,
\label{Ht}
\end{align}
where $J^{}_{L}$ ($J^{}_{R}$) is the bare (i.e., in the absence of the spin-orbit coupling) tunneling amplitude in the left (right) weak link. Here, the spinor $c^{\dagger}_{{\bf k}({\bf p})}=\left[\begin{array}{cc}c^{\dagger}_{{\bf k}({\bf p})\up} &c^{\dagger}_{{\bf k}({\bf p})\down}\end{array}\right]$ creates an electron with wave vector ${\bf k}$ (${\bf p}$) in the left (right) electrode,  and $c^{}_{d}=\left[\begin{array}{c}c^{}_{d\up}\\c^{}_{d\down}\end{array}\right]$ is the annihilation spinor of an electron on the dot. The Hamiltonian describing the entire junction is
\begin{align}
{\cal H}=&\varepsilon^{}_{d}c^{\dagger}_{d}c^{}_{d}+{\cal H}^{}_{\rm tun}(t)\nonumber\\
&+\sum_{\bf k}\varepsilon^{}_{k}c^{\dagger}_{\bf k}c^{}_{\bf k}+\sum_{\bf p}\varepsilon^{}_{p}c^{\dagger}_{\bf p}c^{}_{\bf p}\ .
\label{Ham}
\end{align}
Though the Hamiltonian (\ref{Ham}) is quadratic in the spinor
operators, it turns out that the combination of time dependence and spin-flip terms renders the calculation  (in particular,  the energy fluxes) a rather complicated technical task. We, therefore, assume for simplicity
 that  the  spin-orbit interaction has a negligible effect on the electronic baths (terminals): although the tunneling is time-dependent, the leads stay at thermal equilibrium, and their respective Fermi distributions
\begin{align}
f^{}_{L(R)}(\varepsilon^{}_{k(p)})=&[e^{(\varepsilon^{}_{k(p)}-\mu^{}_{L(R)})/k^{}_{\rm B}T^{}_{L(R)}}+1]^{-1}\ ,
\label{Fermi}
\end{align}
are characterized by the temperatures $T^{}_{L(R)}$ and chemical potentials $\mu^{}_{L(R)}.$

%%%%%%%%%%%%%%%%%%%%%%%%%%%%%%%%%%%%	
%%%%%%%%%%%%%%%%%%%%%%%%%%%%%%%%%%%%

Interesting aspects of the time-dependent particle and energy fluxes associated with the fermionic terminals concern the interrelation between the two. In a static two-terminal device described by a time-independent Hamiltonian,  the energy flux associated with a terminal is just the (energy-resolved) particle flux there, with the extra power of energy in the resulting integration.
We find in  Appendix \ref{expG} that this is not the case when the Hamiltonian is time-dependent (and hence we differ from, e.g.,  Ref. \onlinecite{Bruch2018}).
Another point concerns the separation of the total energy flux into several contributions. We adopt the straightforward picture of assigning an energy flux to each element in the Hamiltonian (\ref{Ham}), see Sec. \ref{fluxes}. The literature, however, offers other options, e.g., conveniently distributing evenly the energy supplied to the junction from the AC field among the dot and the links \cite{Bruch2016}. This again does not agree with our results.

%%%%%%%%%%%%%%%%%%%%%%%%%%%%%%%%%%%%	
%%%%%%%%%%%%%%%%%%%%%%%%%%%%%%%%%%%%

Our paper is constructed as follows. After introducing in Sec. \ref{fluxes} the definitions of the time-dependent particle and energy fluxes flowing in the junction, we present (in Appendix {\ref{Keldysh}) explicit expressions for the Keldysh Green's functions needed in our derivations. Using those we examine in Appendix \ref{sumr} the sum rules obeyed by the fluxes. The complete time dependence of these fluxes is worked out in Appendix \ref{expG}. We use the Keldysh technique of nonequilibrium Green's functions \cite{Jauho1994} combined with Langreth's rules \cite{Lan}, within the approximation termed ``wide-band limit" \cite{Blandin1976, Burrows1994}, which is applied to the fermionic terminals.
Essentially in the wide-band limit the width of the energy band characterizing the electrons' baths
rather than its structure, is the important feature. The main assumption is that the density of states in the terminals, ${\cal N}^{}
_{L(R)}$, is uniform across the band, namely it is assumed that the transport takes place around the common chemical potential of the two terminals. While this assumption does not cause any harm in calculating the particle flux,  it introduces diverging terms when it comes to the energy fluxes; however, we demonstrate that these harmful terms are canceled. These divergencies were noticed in Ref. \onlinecite{Bruch2016} but apparently the authors were not aware of their cancelation.

%%%%%%%%%%%%%%%%%%%%%%%%%%%%%%%%%%%%	
%%%%%%%%%%%%%%%%%%%%%%%%%%%%%%%%%%%%

The complete time dependencies of the fluxes in our junction,  presented in Appendix \ref{Full}, are of use in exploring other properties of the junction, for instance, various noise spectra. Here, however, we study the particle, energy, and heat currents, time-averaged over one driving period of the oscillating field (see Appendix \ref{int}). In addition, we confine ourselves to the regime of weak spin-orbit coupling, assuming that the spin-orbit coupling $k^{}_{\rm so}$ rotates the electron spin as it moves along the weak link (of length $d$) by a small amount, i.e.,  the (dimensionless) quantity $k^{}_{\rm so}d$  is smaller than 1 (see detailed numerical estimates at the end of this section and Ref. \onlinecite{comsoi}). In addition, we
present results expanded to second order in the  frequency $\Omega$, namely, the AC field oscillates very slowly
 (see Appendix \ref{int} and estimates below).  Technically, the expansion turns out to be in powers of the dimensionless $\Omega/(\varepsilon_{d}-\mu)$, multiplied by dimensionless integrals, so that we need $\Omega\ll (\varepsilon_{d}-\mu)$, but it is safer to assume  that $\Omega$ is smaller than all the other energy scales in the problem.

 %%%%%%%%%%%%%%%%%%%%%%%%%%%%%%%%%%%%	
%%%%%%%%%%%%%%%%%%%%%%%%%%%%%%%%%%%%

 We show in Sec. \ref{dotNE}
 that  the junction supports a photovoltaic effect \cite{OEW2020}:
 When mirror symmetry is broken (which happens for $d^{}_{L}\neq d^{}_{R}$, i.e., unequal lengths of the weak links) there appear DC charge and energy currents. The origin of this photovoltaic effect
lies in the different ways inelastic processes modify the reflection of electrons from the junction back into the two terminals, which leads to uncompensated DC transport. The effect, which is even in the AC frequency $\Omega$ and the spin-orbit coupling parameter $k_{\rm so}$, can be detected by measuring the voltage drop or temperature gradient built up due to those  DC currents.
In the present paper, however, we confine ourselves to the more traditional picture of energy and charge currents driven by chemical and temperature differences [see Fig. \ref{sys}(a)], and assume for simplicity weak links of equal length, $d^{}_{L}=d^{}_{R} \equiv \textcolor{blue}{ d}/2$.  This assumption (in addition to the limits of a slow AC frequency and small spin-orbit coupling mentioned above) allows us to display the effect of both the electric-field frequency $\Omega$ and the spin-orbit coupling $k_{\rm so}$ in terms of a single parameter (whose dimensions are energy squared using $\hbar=1$)
 \begin{align}
K^{}_{\rm so}=(\Omega k^{}_{\rm so} d/4)^{2}\ .
\label{KSO}
\end{align}

 %%%%%%%%%%%%%%%%%%%%%%%%%%%%%%%%%%%%	
%%%%%%%%%%%%%%%%%%%%%%%%%%%%%%%%%%%%

Section \ref{dotNQ} presents the charge and heat currents in the linear-response regime, obtained by expanding the Fermi function to linear order in the chemical potential and temperature differences. From these expressions,
one obtains the thermoelectric coefficients, the Seebeck coefficient
$S$, and the electronic thermal conductance $\kappa^{}_e$, Figs. \ref{Sco}  and \ref{Kap}. These figures also contain results of two approximations, explained in
Appendix \ref{LR}. These approximations are limited to low or high temperatures, and they differ significantly from the full numerical calculations. In the figures, $S$ and $\kappa^{}_e$ are drawn versus the dimensionless variable $\beta(\varepsilon^{}_d-\mu)$, where $\beta=1/(k^{}_BT)$, $\mu$ is the common chemical potential of the leads and $\varepsilon^{}_d$ is the single level on the dot. The assumption of a single level requires that $\beta(\varepsilon^{}_{d}-\mu)\gg 1$, and in that region both $S$ and $\kappa^{}_e$ are enhanced by the spin-orbit interaction. There are intermediate regions where one (or both) decreases.

%%%%%%%%%%%%%%%%%%%%%%%%%%%%%%%%%%%%	
%%%%%%%%%%%%%%%%%%%%%%%%%%%%%%%%%%%%

Interestingly,  we find that the averaged power supplied by the AC field appears in the heat currents emerging from the left and right terminals [see Fig. \ref{sys}(b) and Sec. \ref{average}], while not affecting the energy fluxes in the weak links (which vanish upon being averaged over a period of the AC field).  In this respect, the results of our analysis differ from previous ones, e.g., Refs. \onlinecite{Ludovico2016,Ryo2022}. This difference seems to be crucial when it comes to the precise definitions of efficiencies, as discussed in Sec. \ref{effs}, where the conditions on the junction to operate as a heat engine or as a heat pump are analyzed. The power supply which forms a part of the heat flux involved with each terminal is (to leading order)  not affected by $\beta_{L}-\beta_{R}$ or $\mu^{}_{R}-\mu^{}_{L} $, i.e., by the thermoelectric driving forces (see Fig. \ref{sys}). It follows that the heat flux of each terminal includes a term independent of the driving forces. This power term modifies the working conditions of the junction and confines the possible range of the driving forces.

 %%%%%%%%%%%%%%%%%%%%%%%%%%%%%%%%%%%%	
%%%%%%%%%%%%%%%%%%%%%%%%%%%%%%%%%%%%

Estimates for the magnitude of the Rashba parameter $k^{}_{\rm so}d$ which determines the Aharonov-Casher phase factors, Eqs. (\ref{acL}),
can be extracted from the literature. For instance, for a spin-orbit active weak link in the form of an InAs nanowire, one
may adopt the value $k_{\rm so}^{}=[100\ {\rm nm}]^{-1}$ measured in Ref. \onlinecite{Scherubl2016}.
A wire length of $ d=100$ nm would then give $k^{}_{\rm so} d=1$. For a  distance of 200 nm between the side gates supplying the electric field
\cite{Scherubl2016}, a microwave-generated amplitude  of 1 V on the side gate (see Fig. \ref{sys}(a))
would produce a transverse electric field amplitude of 50 kV/cm in the wires, corresponding to a
Rashba parameter $\alpha^{}_{R}=\hbar^{2}k^{}_{\rm so}/m^{\ast}\approx$ 50 meV \AA $\ $ \cite{Luo2023} and, using for the effective mass $m^{\ast}_{}=0.023m_{\rm e}^{}$, a Rashba coupling $k^{}_{\rm so}\approx 20\times 10^{-3}\ ({\rm nm})^{-1}$
appears in the weak links. With $d^{}_{L}\approx d^{}_{R}\approx 250$ nm one finds
$k^{}_{\rm so} d/2\approx 0.5$. We choose the microwave frequency to be $2\pi\times 100$ GHz so that $\hbar\Omega\approx 0.4 $ meV  which is smaller than the energy level on the dot
 \cite{Kouwenhoven1997},
$\varepsilon^{}_{d}-\mu\sim 1$ meV [$\mu=(\mu^{}_{L}+\mu^{}_{R})/2$ is the common chemical potential of the junction],
so that $ K^{}_{\rm so}/(\varepsilon^{}_{d}-\mu)^{2}\sim 1/16$.

%%%%%%%%%%%%%%%%%%%%%%%%%%%%%%%%%%%%	
%%%%%%%%%%%%%%%%%%%%%%%%%%%%%%%%%%%%

%%%%%%%%%%%%%%%%%%%%%%%%%%%%%%%%%%%%	
%%%%%%%%%%%%%%%%%%%%%%%%%%%%%%%%%%%%

\section{Particle and energy fluxes}
\label{fluxes}
%%%%%%%%%%%%%%%%%%%%%%%%%%%%%%%%%%%%
%%%%%%%%%%%%%%%%%%%%%%%%%%%%%%%%%%%%%

The time-dependent  particle flux associated with the left electrons' bath is \cite{comcur}
 \begin{align}
\dot{N}^{}_{L}(t)\equiv
&\frac{d}{dt} \sum_{\bf k}\langle c^{\dagger}_{\bf k}(t)c^{}_{\bf k}(t)\rangle\ ,
 \label{dNL}
\end{align}
where the angular brackets indicate a quantum average. Exploiting the Hamiltonian (\ref{Ham}) to calculate the time derivative, one finds
 \begin{align}
\dot{N}^{}_{L}(t)
 =J^{}_{L}\sum_{\bf k}{\rm Tr}\{e^{-i\varphi^{}_{L}(t)}G^{<}_{{\bf k}d}(t,t)-G^{<}_{d{\bf k}}(t,t)e^{i\varphi^{}_{L}(t)} \}\ ,
 \label{dNLG}
\end{align}
where $G^{<}_{{\bf k}d}$ is the lesser Green's function, e.g.,
\begin{align}
G^{<}_{{\bf k}\sigma,d\sigma'}(t,t')=i\langle c^{\dagger}_{d\sigma'}(t')c^{}_{{\bf k}\sigma}(t)\rangle\ ,
\label{Gkd}
\end{align}
with an analogous expression for $G^{<}_{d{\bf k}}$ \cite{Jauho2003, Odashima2017}. (The Green's functions are matrices in spin space.)
 The particle flux associated with the right bath is obtained from Eq.
(\ref{dNLG})  by replacing $L$ with $R$ and ${\bf k}$ with ${\bf p}$. This scheme (of interchanging $L$ with $R$ and ${\bf k}$ with ${\bf p}$) pertains to all other fluxes and Green's functions encountered below, whose derivation is given in Appendix \ref{Keldysh}.

%%%%%%%%%%%%%%%%%%%%%%%%%%%%%%%%%%%%	
%%%%%%%%%%%%%%%%%%%%%%%%%%%%%%%%%%%%

The particle flux associated with the dot is
\begin{align}
\dot{N}_{d}^{}=-i{\rm Tr}\{dG^{<}_{dd}(t,t)/dt\}\ .
\label{Id}
\end{align}
We show in Appendix \ref{sumr} that
\begin{align}
\dot{N}^{}_{L}(t)+\dot{N}^{}_{R}(t)+\dot{N}^{}_{d}(t)=0\ ,
\label{cc}
\end{align}
i.e., charge is conserved.

%%%%%%%%%%%%%%%%%%%%%%%%%%%%%%%%%%%%	
%%%%%%%%%%%%%%%%%%%%%%%%%%%%%%%%%%%%

The energy flux associated with the dot is obviously
\begin{align}
\dot{E}^{}_{d}(t)=\varepsilon^{}_{d}\dot{N}^{}_{d}(t)\ ,
\label{dEd}
\end{align}
where the particle flux the dot, $\dot{N}^{}_{d}(t)$,  is given in Eq. (\ref{Id}).
The energy flux associated with the left reservoir, $\dot{E}^{}_{L}(t)$,
\begin{align}
\dot{E}^{}_{L}(t)\equiv
&\frac{d}{dt} \sum_{\bf k}\varepsilon^{}_{k}\langle c^{\dagger}_{\bf k}(t)c^{}_{\bf k}(t)\rangle\ ,
\label{dEL}
\end{align}
is derived along the same steps used to obtain $\dot{N}_{L}(t)$, with the only (and very crucial) difference that the sum over the wave vector ${\bf k}$ contains the energy $\varepsilon^{}_{k}$,
\begin{align}
\dot{E}^{}_{L}(t)
 =J^{}_{L}\sum_{\bf k}\varepsilon^{}_{k}{\rm Tr}\{e^{-i\varphi^{}_{L}(t)}G^{<}_{{\bf k}d}(t,t)-G^{<}_{d{\bf k}}(t,t)e^{i\varphi^{}_{L}(t)} \}\ .
 \label{dELG}
\end{align}
Finally, the energy flux associated with the left weak link, $\dot{E}^{}_{L,{\rm tun}}(t)$, is
\begin{align}
\dot{E}^{}_{L,{\rm tun}}(t)\equiv J^{}_{L}
&\frac{d}{dt} \sum_{\bf k}\langle c^{\dagger}_{\bf k}(t)\exp[i\varphi^{}_{L}(t)]c^{}_{d}(t)\rangle+{\rm H.c.}\ ,
\label {dELtun}
\end{align}
i.e., [see Eq. (\ref{Gkd})]
\begin{align}
\dot{E}^{}_{L,{\rm tun}}(t)=-iJ^{}_{L}\sum_{\bf k}{\rm Tr}\{\frac{d}{dt}[G^{<}_{d{\bf k}}(t,t) e^{i\varphi^{}_{L}(t)}+
e^{-i\varphi^{}_{L}(t)}G^{<}_{{\bf k}d}(t,t)]
\}\ .
 \label{dEtun}
\end{align}
We consider in Appendix \ref{sumr} the sum of all energy fluxes in our junction. In particular, we show that this sum vanishes (i.e., energy is conserved) provided the spin-orbit coupling is static. The presence of the time-dependent Aharonov-Casher factors implies that energy is not conserved in the junction:  extra energy is supplied {\it to all parts of the junction} by the AC field via the spin-orbit terms.

 %%%%%%%%%%%%%%%%%%%%%%%%%%%%%%%%%%%%	
%%%%%%%%%%%%%%%%%%%%%%%%%%%%%%%%%%%%

\section{Averaged thermal-electric transport}
\label{average}

\subsection{Averaged particle and energy fluxes}
\label{dotNE}

As stated above,  we examine the thermoelectric properties of our junction by employing fluxes averaged over a single oscillation of the microwave field. The averaged particle flux associated with the left terminal ~in the weak Rashba coupling and small $\Omega$ limits is [see Appendix \ref{expG}]
\begin{align}
\overline{\dot{N}}^{}_{L}\equiv\frac{\Omega}{2\pi}\int _{0}^{\frac{2\pi}{\Omega}}&dt\dot{N}^{}_{L}(t)=8\Gamma^{}_{L}\Gamma^{}_{R}\int\frac{d\omega}{2\pi}\Big ([f^{}_{R}(\omega)-f^{}_{L}(\omega)]D(\omega)\nonumber\\
&+\frac{(k^{}_{\rm so}\Omega)^{2}}{4}[d^{2}_{R}f^{}_{R}(\omega)-d^{2}_{L}f^{}_{L}(\omega)]D''(\omega)\Big)\ ,
\label{ILav}
\end{align}
where
\begin{align}
D(\omega)=[(\omega-\varepsilon^{}_{d})^{2}+\Gamma^{2}]^{-1}\ .
\label{D}
\end{align}
Here, $\Gamma$ is the width of the Breit-Wigner resonance formed on the dot due to the coupling with the terminals.
Adding to Eq. (\ref{ILav}) the period average of $\dot{N}^{}_{R}(t)$ gives zero. It then follows that the period average of the particle flux associated with the dot and the corresponding energy flux [Eq. (\ref{dEd})],  vanish. One notes that particle current is flowing even in the absence of a bias or a temperature gradient, i.e., when $f_{L}(\omega)=f_{R}(\omega)$  \cite{OEW2020}. The flow direction is fixed by the length ratio of the weak links. Upon assuming that the lengths of the weak links are identical, the second term in the circular brackets becomes $K_{\rm so}[f^{}_{R}(\omega)-f^{}_{L}(\omega)]D''(\omega)$, with $K_{\rm so}$ given in Eq. (\ref{KSO}).

The period average of the energy flux associated with the left electrons' bath is obtained by
adding together  Eqs. (\ref{IELa}) and (\ref{IELb}). It consists of two parts, i.e.,
\begin{align}
\overline{\dot{E}}^{}_{L}\equiv&\frac{\Omega}{2\pi}\int^{\frac{2\pi}{\Omega}}_{0}dt\dot{E}^{}_{L}(t)=P^{}_{L}\nonumber\\
&
+8\Gamma^{}_{L}\Gamma^{}_{R}\int\frac{d\omega}{2\pi}\Big ([f^{}_{R}(\omega)-f^{}_{L}(\omega)]\omega D(\omega)\nonumber\\
&+\frac{(k^{}_{\rm so}\Omega)^{2}}{4}[f^{}_{R}(\omega)d^{2}_{R}-f^{}_{L}(\omega)d^{2}_{L}]\omega D''(\omega)\Big )\ .
\label{ELav}
\end{align}
The {\it second} term on the right-hand-side is of the ubiquitous form: it consists of the same frequency integral as the one that determines the particle flux, Eq. (\ref{ILav}),  with the extra frequency power in the integrand. On the other hand, the first term in Eq. (\ref{ELav}) is the flux of energy supplied to the left lead by the AC field creating the spin-orbit coupling,
\begin{align}
P^{}_{L}=4\Gamma^{}_{L}\Gamma^{}_{R}(k^{}_{\rm so}\Omega)^{2}d^{}_{R}(d^{}_{L}+d^{}_{R})\int\frac{d\omega}{2\pi}[-f'_{R}(\omega)]D(\omega)\ ,
\label{PL0}
\end{align}
which is positive.
The extra energy supplied to the left electrons' bath, {\it even when averaged over the period}, is crucial in forming the thermoelectric properties of the junction.

 %%%%%%%%%%%%%%%%%%%%%%%%%%%%%%%%%%%%	
%%%%%%%%%%%%%%%%%%%%%%%%%%%%%%%%%%%%

Adding  to Eq. (\ref{ELav}) the corresponding contribution from the right fermionic terminal one finds
\begin{align}\label{EA}
\overline{\dot{E}}^{}_{L}+\overline{\dot{E}}^{}_{R}=&
4\Gamma^{}_{L}\Gamma^{}_{R}(k^{}_{\rm so}\Omega)^{2}(d^{}_{L}+d^{}_{R})\nonumber\\
&\times\int\frac{d\omega}{2\pi}[d^{}_{R}f^{}_{R}(\omega)+d^{}_{L}f^{}_{L}(\omega)]D'(\omega)\ .
\end{align}
This is indeed the period average of the sum of all energy fluxes, see Eq. (\ref{sumE}).
Put differently, the period average of the energy flux associated with each of the weak links vanishes, i.e.,
$\overline{\dot{E}}^{}_{L(R),{\rm tun}}(t)=0$.

 %%%%%%%%%%%%%%%%%%%%%%%%%%%%%%%%%%%%	
%%%%%%%%%%%%%%%%%%%%%%%%%%%%%%%%%%%%

 It is interesting to consider the fluxes when the junction is not biased, i.e., $\beta^{}_{L}=\beta^{}_{R}$ and $\mu^{}_{L}=\mu^{}_{R}$. In that case,
 particle current, Eq. (\ref{ILav}), flows when $d^{}_{L}\neq d^{}_{R}$. At zero temperature,  its magnitude is determined  by  \cite{OEW2020}
 \begin{align}
 \int_{-\infty}^{\mu} d\omega D''(\omega)=D'(\omega)\Big|^{\mu}_{-\infty}=\frac{2(\varepsilon^{}_{d}-\mu)}{[(\varepsilon^{}_{d}-\mu)^{2}+\Gamma^{2}]^{2}} \ .
 \end{align}
 The corresponding quantity in the heat flux is
 \begin{align}
 -\int_{-\infty}^{\mu} d\omega (\mu-\omega)D''(\omega)  =-\int_{-\infty}^{\mu} d\omega D'(\omega) =-D(\mu) \ .
 \end{align}
 Adding this expression to $P_{L}$ (calculated at zero temperature), we find that the total is half the power supplied to the junction.

%%%%%%%%%%%%%%%%%%%%%%%%%%%%%%%%%%%%	
%%%%%%%%%%%%%%%%%%%%%%%%%%%%%%%%%%%%

In the following, we consider a junction biased by a chemical potential difference, a temperature difference, or both.
By the Clausius relation, the average entropy production,  $\overline{\dot{\cal S}}$,  in a two-terminal  junction reads
\begin{align}
\overline{\dot{S}}^{}_{}=[\overline{\dot{E}}^{}_{L}-\mu^{}_{L}\overline{\dot{N}}^{}_{L}]/T^{}_{L}+
[\overline{\dot{E}}^{}_{R}-\mu^{}_{R}\overline{\dot{N}}^{}_{R}]/T^{}_{R}\ .
\label{entp0}
\end{align}
Written in terms of heat fluxes
\begin{align}
\overline{\dot{Q}}^{}_{L,R}=\overline{\dot{E}}^{}_{L,R}-\mu^{}_{L,R}\overline{\dot{N}}^{}_{L,R}\ ,
\label{dQ0}
\end{align}
the Clausius relation becomes
\begin{align}
T\overline{\dot{\cal S}}=\overline{\dot{Q}}^{}_{L}\frac{\beta^{}_{L}-\beta^{}_{R}}{\beta}+\overline{\dot{N}}^{}_{L}(\mu^{}_{R}-\mu^{}_{L})+P\ ,
\label{TS}
\end{align}
where
$P=P^{}_{L}+P^{}_{R}=\overline{\dot{E}}^{}_{L}+\overline{\dot{E}}^{}_{R}$ is the total averaged power supplied by the AC source [which is positive, see Eq. (\ref{PL0})]}.

%%%%%%%%%%%%%%%%%%%%%%%%%%%%%%%%%%%%	
%%%%%%%%%%%%%%%%%%%%%%%%%%%%%%%%%%%%

\subsection{Linear-response, Onsager's relations }
\label{dotNQ}

Thermoelectric properties are intrinsically linked with the system's coupling to the external world: the relevant coefficients are defined in response to chemical potential and temperature differences (in linear order), which implies the expansion of $f_{L}(\omega)\neq f^{}_{R}(\omega)$. Below we assume for convenience that the weak links are of equal lengths \cite{Com1}, i.e.,  $d_{L}=d_{R}=\textcolor{blue}{ d}/2$, and consequently introduce into the expressions the spin-orbit coupling in terms of $K_{\rm so}$, Eq. (\ref{KSO}).

%%%%%%%%%%%%%%%%%%%%%%%%%%%%%%%%%%%%	
%%%%%%%%%%%%%%%%%%%%%%%%%%%%%%%%%%%%

In linear response,
\begin{align}
f^{}_{L}(\omega)\approx f(\omega)+\frac{f'(\omega)}{2}[\mu^{}_{R}-\mu^{}_{L}+(\omega-\mu)\frac{\beta^{}_{L}-\beta^{}_{R}}{\beta}]\ ,
\label{LIN}
\end{align}
where $D(\omega)$  is defined in Eq. (\ref{D}),
\begin{align}
\mu=(\mu^{}_{L}+\mu^{}_{R})/2\ ,\ \ \ \beta=(\beta^{}_{L}+\beta^{}_{R})/2\ ,
\end{align}
and $f(\omega)=[\exp[\beta(\omega-\mu)]+1]^{-1}$.
The particle flux, Eq. (\ref{ILav}), then reads
\begin{align}
\overline{\dot{N}}^{}_{L}=
I^{}_{0}(\mu^{}_{R}-\mu^{}_{L})+
I^{}_{1}\frac{\beta^{}_{L}-\beta^{}_{R}}{\beta}\ ,
\label{NL}
\end{align}
where
\begin{align}
I^{}_{n}=I^{(0)}_{n}+K^{}_{\rm so}I^{(2)}_{n}\ , \ \ n=0\ , 1\ , 2\ ,
\label{In}
\end{align}
with
\begin{align}
I^{(\ell)}_{n}%=\int\frac{d\omega}{2\pi}[-f'(\omega)](\omega-\mu)^{n}F_{}^{(\ell)}(\omega)
=\frac{4\Gamma^{}_{L}\Gamma^{}_{R}}{\pi}\int d\omega
[-f'(\omega)](\omega-\mu)^{n}_{}D_{}^{(\ell)}(\omega)\ .
\label{sch}
\end{align}
Here $D_{}^{(1)}(\omega)=D'(\omega)$, and $D_{}^{(2)}(\omega)=D''(\omega)$. We present in Appendix \ref{LR} two approximations for the integrals (\ref{In}) valid at relatively high temperatures (valid for $\beta\Gamma<<1$), and low temperatures [$\beta(\varepsilon^{}_{d}-\mu)>1$].

%%%%%%%%%%%%%%%%%%%%%%%%%%%%%%%%%%%%	
%%%%%%%%%%%%%%%%%%%%%%%%%%%%%%%%%%%%
%\begin{figure}
%\includegraphics[width=0.43\textwidth]{dS-1.pdf}
%\caption{(color online) The contribution of the spin-orbit coupling to the Seebeck coefficient, see Eq. (\ref{S}). The solid (blue) curve is obtained by solving numerically the integrals $I^{}_{n}$, for $\beta\Gamma=0.2$ and $K^{}_{\rm so}/(\varepsilon^{}_{d}-\mu)^{2}=1/16$ \cite{OEW2020}. The other two curves are the approximations of the integrals discussed in Appendix \ref{LR}: dashed (red) line: low-temperature approximation, dashed-dotted (red) line: high-temperature approximation. Here, $S^{}_{0}=I^{(0)}_{1}/I^{(0)}_{0}$ is the Seebeck coefficient for $K^{}_{\rm so}=0$.}
%\label{delS}
%\end{figure}
%%%%%%%%%%%%%%%%%%%%%%%%%%%%%%%%%%%%	
%%%%%%%%%%%%%%%%%%%%%%%%%%%%%%%%%%%%

%%%%%%%%%%%%%%%%%%%%%%%%%%%%%%%%%%%%	
%%%%%%%%%%%%%%%%%%%%%%%%%%%%%%%%%%%%
\begin{figure}
\includegraphics[width=0.43\textwidth]{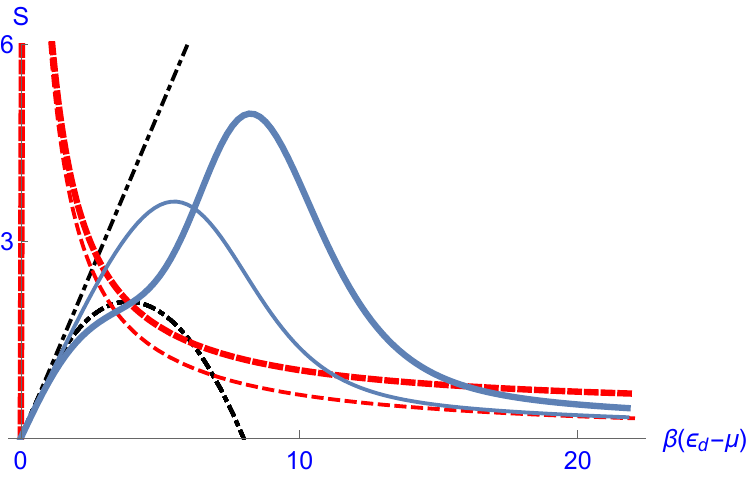}
\caption{(color online)
The Seebeck coefficient (in units of $k^{}_B/|e|$), Eq. (\ref{S}). The
solid (blue) curves are obtained by calculating the integrals $I^{(\ell)}_{n}$ numerically,
for $\beta\Gamma=0.2$ and $K^{}_{\rm so}/(\varepsilon^{}_{d}-\mu)^{2}=1/16$ (thick lines) or $K^{}_{\rm so}=0$ (thin lines)
\cite{OEW2020}. In each case, the other two curves are the approximations of the integrals discussed in Appendix \ref{LR}: dashed (red) line: low-temperature approximation, dashed-dotted (black) line: high-temperature approximation.}
\label{Sco}
\end{figure}
%%%%%%%%%%%%%%%%%%%%%%%%%%%%%%%%%%%%	
%%%%%%%%%%%%%%%%%%%%%%%%%%%%%%%%%%%%

%%%%%%%%%%%%%%%%%%%%%%%%%%%%%%%%%%%%	
%%%%%%%%%%%%%%%%%%%%%%%%%%%%%%%%%%%%

The heat flux associated with the left terminal [Eq. (\ref{dQ0})], %is
%\begin{align}
%\overline{\dot{Q}}^{}_{L}= \overline{\dot{E}}^{}_{L}-\mu^{}_{L}\overline{\dot{N}}^{}_{L}= \overline{\dot{E}}^{}_{L}-\mu\overline{\dot{N}}^{}_{L}-\frac{\mu^{}_{L}-\mu^{}_{R}}{2}\overline{\dot{N}}^{}_{L}\ .
%\end{align}
to linear order in the chemical-potential and temperature differences, is % it becomes [{\it cf.} Eq. (\ref{ELav})]
\begin{align}
\overline{\dot{Q}}^{}_{L}&=P^{}_{L}+
I^{}_{1}(\mu^{}_{R}-\mu^{}_{L})
+I^{}_{2}\frac{\beta^{}_{L}-\beta^{}_{R}}{\beta}\ ,
\label{QL}
\end{align}
where
\begin{align}
P^{}_{L}=&\frac{16\Gamma^{}_{L}\Gamma^{}_{R}}{\pi}K^{}_{\rm so}\int d\omega[-f'^{}_{R}(\omega)]D(\omega)\approx 4K^{}_{\rm so}I^{(0)}_{0}\ .
\label{PL}
\end{align}
Comparing Eq. (\ref{NL}) with Eq. (\ref{QL}) shows that Onsager's relations are fulfilled (in terms of the response to the chemical potential and temperature differences). However, the heat fluxes also include the power supplied by the AC field.
In this sense, our system differs from others studied in the literature, e.g., Refs. \onlinecite{Ludovico2016,Ryo2022}.

%%%%%%%%%%%%%%%%%%%%%%%%%%%%%%%%%%%%	
%%%%%%%%%%%%%%%%%%%%%%%%%%%%%%%%%%%%

Equations  (\ref{NL}) and (\ref{QL}) yield the thermoelectric coefficients of the junction. The ones crucial for the thermoelectric performance are the Seebeck coefficient and the electronic thermal conductance.
The Seebeck coefficient, S,  is $I^{}_{1}/I^{}_{0}$, in units of $k^{}_{\rm B}/e$ ($I^{}_{0}$ is the electrical conductance, in units of the quantum conductance). To linear order in $K^{}_{\rm so}$ (quadratic in $\Omega$)
\begin{align}
S\propto\frac{I^{}_{1}}{I^{}_{0}}=\frac{I^{(0)}_{1}}{I^{(0)}_{0}}\frac{1+K^{}_{\rm so}\frac{I^{(2)}_{1}}{I^{(0)}_{1}}}{1+K^{}_{\rm so}\frac{I^{(2)}_{0}}{I^{(0)}_{0}}}\approx \frac{I^{(0)}_{1}}{I^{(0)}_{0}}\Big (1+K^{}_{\rm so}\Big [\frac{I^{(2)}_{1}}{I^{(0)}_{1}}-\frac{I^{(2)}_{0}}{I^{(0)}_{0}}\Big]\Big )
\ .
\label{S}
\end{align}
Figure \ref{Sco} presents $S$, Eq, (\ref{S}), obtained by computing numerically the integrals (\ref{sch}), together
with its two approximations [see  Appendix \ref{LR}]. The thin curves display the same quantities in the absence of the spin-orbit coupling, i.e., for $K_{\rm so}=0$' One observes that
for $\beta(\varepsilon^{}_{d}-\mu)>6.5$,
$S$ is increased (considerably) by $K_{\rm so}$. This increase is advantageous for good thermoelectric performance.

%%%%%%%%%%%%%%%%%%%%%%%%%%%%%%%%%%%%	
%%%%%%%%%%%%%%%%%%%%%%%%%%%%%%%%%%%%

\begin{figure}
\includegraphics[width=0.43\textwidth]{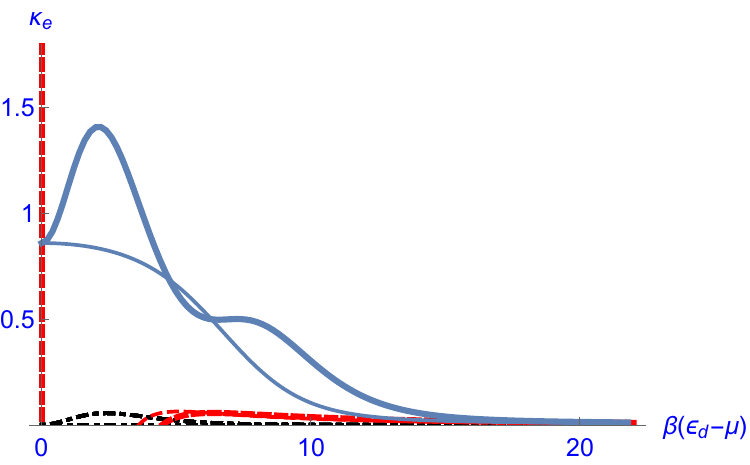}
\caption{(color online) Same as Fig. \ref{Sco}, for $\kappa^{}_{\rm e}$, the electronic thermal conductance (divided by the amplitude $4\Gamma^{}_{L}\Gamma^{}_{R}/\pi$). }
\label{Kap}
\end{figure}

The electronic thermal conductance is
\begin{align}
&\kappa^{}_{\rm e}\propto I^{}_{2}-\frac{I^{2}_{1}}{I^{}_{0}}\approx \kappa^{}_{\rm e0}
+K^{}_{\rm so}\Big (I^{(2)}_{2}-\frac{2I^{(2)}_{1}I^{(0)}_{1}}{I^{(0)}_{0}}+I^{(2)}_{0}[\frac{I^{(0)}_{1}}{I^{(0)}_{0}}]^{2}_{}\Big )\ ,
\label{kapa}
\end{align}
(in units of absolute temperature times the quantum conductance divided by the electron charge), where $\kappa^{}_{\rm e0}=I^{(0)}_{2}-[I^{0}_{1}]^{2}_{}/I^{(0)}_{0}$.
Unfortunately,
the contribution of the spin-orbit coupling to the thermal conductance on most of the range is positive, see Fig. \ref{Kap}. The plot shows a negative contribution to $\kappa^{}_{\rm e}$, but only for a narrow region of $\beta(\varepsilon^{}_{d}-\mu)$ values, where $S$ is also decreased. Interestingly, for very large values of $\beta(\varepsilon^{}_{d}-\mu)$ (very low temperatures) the relative changes in $S$ and in $\kappa^{}_{\rm e}$ approach the same limit $3/8 $ as is also given by the low-temperature approximation (see more details in Appendix \ref{LR}).

%%%%%%%%%%%%%%%%%%%%%%%%%%%%%%%%%%%%	
%%%%%%%%%%%%%%%%%%%%%%%%%%%%%%%%%%%%
%\begin{figure}
%\includegraphics[width=0.43\textwidth]{dK-1.pdf}
%\caption{(color online) The contribution of the spin-orbit coupling to the electronic thermal conductance, see Eq. (\ref{kapa}). The solid (blue) curve is obtained by solving numerically the integrals $I^{}_{n}$, for $\beta\Gamma=0.2$ and $K^{}_{\rm so}/(\varepsilon^{}_{d}-\mu)^{2}=1/16$ \cite{OEW2020}. The other dashed-line curves are the approximations of the integrals at low temperatures.  }
%\label{delKap}
%\end{figure}
%%%%%%%%%%%%%%%%%%%%%%%%%%%%%%%%%%%%	
%%%%%%%%%%%%%%%%%%%%%%%%%%%%%%%%%%%%

%%%%%%%%%%%%%%%%%%%%%%%%%%%%%%%%%%%%	
%%%%%%%%%%%%%%%%%%%%%%%%%%%%%%%%%%%%
%%%%%%%%%%%%%%%%%%%%%%%%%%%%%%%%%%%%	
%%%%%%%%%%%%%%%%%%%%%%%%%%%%%%%%%%%%

%%%%%%%%%%%%%%%%%%%%%%%%%%%%%%%%%%%%	
%%%%%%%%%%%%%%%%%%%%%%%%%%%%%%%%%%%%
\subsection{Efficiencies}
\label{effs}

The possible thermoelectric efficiencies of the junction can be defined in various ways.   For example, Ref. \onlinecite{Ludovico2016} (see also Ref. \onlinecite{Ryo2022}) defines the efficiency of a junction working as a heat engine at zero bias ($\mu^{}_{L}=\mu^{}_{R}$), by assuming that heat leaving the left terminal, $-\overline{\dot{Q}}^{}_{L}$ (assuming that $\beta^{}_{L}<\beta^{}_{R}$) enables the electrons
to perform work,  $-P$,  on the AC source, leading to $\eta^{}_{\rm he}=P/\overline{\dot{Q}}^{}_{L}$. Here we
adopt the more ``traditional" approach, namely  the junction supplies electric power,
\begin{align}
P^{}_{\rm out}=(\mu^{}_{R}-\mu^{}_{L})[-I^{}_{0}(\mu^{}_{R}-\mu^{}_{L})+
I^{}_{1}\frac{\beta^{}_{R}-\beta^{}_{L}}{\beta}]\ ,
\end{align}
with the working condition
\begin{align}
-I^{}_{0}(\mu^{}_{R}-\mu^{}_{L})+
I^{}_{1}\frac{\beta^{}_{R}-\beta^{}_{L}}{\beta}>0\ ,
\label{wc1}
\end{align}
(when $-\overline{\dot{N}}^{}_{L}>0$ and $\mu^{}_{R}>\mu^{}_{L}$), on the expense of the heat supplied, $-\overline{\dot{Q}}^{}_{L}$ which should be positive,
\begin{align}
-I^{}_{1}(\mu^{}_{R}-\mu^{}_{L})+I^{}_{2}\frac{\beta^{}_{R}-\beta^{}_{L}}{\beta}>P^{}_{L}\ .
\label{wc2}
\end{align}
Hence
\begin{align}
\eta^{}_{\rm he}=\frac{(\mu^{}_{R}-\mu^{}_{L})[-I^{}_{0}(\mu^{}_{R}-\mu^{}_{L})+
I^{}_{1}\frac{\beta^{}_{R}-\beta^{}_{L}}{\beta}]}{-I^{}_{1}(\mu^{}_{R}-\mu^{}_{L})+I^{}_{2}\frac{\beta^{}_{R}-\beta^{}_{L}}{\beta}-P^{}_{L}}\ ,
\label{eta2}
\end{align}
provided that the working conditions, Eqs. (\ref{wc1}) and (\ref{wc2}),  are obeyed.

%%%%%%%%%%%%%%%%%%%%%%%%%%%%%%%%%%%%	
%%%%%%%%%%%%%%%%%%%%%%%%%%%%%%%%%%%%

The junction will work as a heat pump (i.e., a thermoelectric refrigerator) when it cools the left terminal, $\beta^{}_{R}<\beta^{}_{L}$,
on the expense of Joule power supplied to it
\begin{align}
(\mu^{}_{L}-\mu^{}_{R})[I^{}_{0}(\mu^{}_{L}-\mu^{}_{R})-
I^{}_{1}\frac{\beta^{}_{L}-\beta^{}_{R}}{\beta}]>0\ .
\label{wc3}
\end{align}
The working condition ensuring that the  cooling power is positive, reads
 \begin{align}
%-\overline{\dot{Q}}^{(0)}_{L}&=
I^{}_{1}(\mu^{}_{L}-\mu^{}_{R})
-
I^{}_{2}\frac{\beta^{}_{L}-\beta^{}_{R}}{\beta}>P^{}_{L}\ .
\label{wc4}
\end{align}
The corresponding $\eta^{}_{\rm hp}$ (sometimes called coefficient of performance) is
\begin{align}
\eta^{}_{\rm hp}&=\frac{I^{}_{1}(\mu^{}_{L}-\mu^{}_{R})-
I^{}_{2}\frac{\beta^{}_{L}-\beta^{}_{R}}{\beta}-P^{}_{L}}{(\mu^{}_{L}-\mu^{}_{R})[I^{}_{0}(\mu^{}_{L}-\mu^{}_{R})-
I^{}_{1}\frac{\beta^{}_{L}-\beta^{}_{R}}{\beta}]}\ ,
\end{align}
with $\eta^{}_{\rm he}=-1/\eta^{}_{\rm hp}$. The Carnot limit on $\eta^{}_{\rm he}$ is $(\beta^{}_{R}-\beta^{}_{L})/\beta$
while that on $\eta^{}_{\rm hp}$ is $\beta/(\beta^{}_{L}-\beta^{}_{R})$. In terms of the entropy production $T\overline{\dot{\cal S}}$ [Eq. (\ref{TS})], we find
\begin{align}
\eta^{}_{\rm he}&=\frac{\beta^{}_{R}-\beta^{}_{L}}{\beta}+\frac{-T\overline{\dot{\cal S}}+P}{-I^{}_{1}(\mu^{}_{R}-\mu^{}_{L})+I^{}_{2}\frac{\beta^{}_{R}-\beta^{}_{L}}{\beta}-P^{}_{L}}\ ,\nonumber\\
\eta^{}_{\rm hp}&=\frac{\beta}{\beta^{}_{L}-\beta^{}_{R}}\Big (1+\frac{-T\overline{\dot{\cal S}}+P}{(\mu^{}_{L}-\mu^{}_{R})[I^{}_{0}(\mu^{}_{L}-\mu^{}_{R})-
I^{}_{1}\frac{\beta^{}_{L}-\beta^{}_{R}}{\beta}]}\Big )\ .
\label{2eta}
\end{align}
Inspecting Eqs. (\ref{2eta}), one may conclude that when the entropy production vanishes, both efficiencies exceed the Carnot limit. However, this does not happen.  The entropy production Eq. (\ref{TS}) can be written in the form
\begin{align}
T\overline{\dot{\cal S}}&=P+\frac{\beta^{}_{L}-\beta^{}_{R}}{\beta}P^{}_{L}\nonumber\\
&+I^{}_{0}\Big (\frac{\beta^{}_{L}-\beta^{}_{R}}{\beta}\Big )^{2}_{}\Big [\frac{I^{}_{0}I^{}_{2}-I^{2}_{1}}{I^{2}_{0}}+\Big (\frac{\beta(\mu^{}_{R}-\mu^{}_{L})}{\beta^{}_{L}-\beta^{}_{R}}+\frac{I^{}_{1}}{I^{}_{0}}\Big )^{2}_{}\Big ]\ .
\end{align}
In a symmetric junction,  $P_{L}^{}=P^{}_{R}$ [since $d_{L}^{}=d^{}_{R}$, see Eq. (\ref{PL})], and then $P+P^{}_{L}(\beta^{}_{L}-\beta^{}_{R})/\beta=P\beta^{}_{L}/\beta\approx P$, leading to a negative value for $-T\overline{\dot{\cal S}}+P$.
[Recall that $[I^{}_{0}I^{}_{2}-I^{2}_{1}]/I^{}_{0}$, the electronic thermal conductance -- which is proportional to the inverse of the figure of merit -- is positive.]

%%%%%%%%%%%%%%%%%%%%%%%%%%%%%%%%%%%%	
%%%%%%%%%%%%%%%%%%%%%%%%%%%%%%%%%%%%
\section{Summary}
\label{sum}

The possibility of enhancing the thermoelectric functionality by employing the spin degree of freedom via the spin-orbit interaction has appeared recently in the literature, mainly in connection with two-dimensional layers (see, for instance, Refs. \onlinecite{Islam2012, Xiao2016, Yuan2018, Tang2022}). These works rely on the intrinsic Rashba coupling of the system and generally find that the thermoelectric performance is improved. Thus, for example, Ref. \onlinecite{Xiao2016} reports an enhancement connected with the topological band-crossing point revealed once elastic scattering is treated beyond the relaxation-time approximation. (Recall that in Appendix \ref{LR} we also obtain that within the high-temperature approximation, the thermoelectric performance has deteriorated.) Yuan {\it et al. } \cite{Yuan2018} discuss theoretically the thermoelectric performance of bismuth antimony sheets, and find a figure of merit that at room temperature (!) is doubled compared to the spin-degenerate case.

%%%%%%%%%%%%%%%%%%%%%%%%%%%%%%%%%%%%	
%%%%%%%%%%%%%%%%%%%%%%%%%%%%%%%%%%%%

The Rashba coupling in the junction we study is induced by an AC field and modifies the thermoelectric coefficients through a prefactor [$K_{\rm so}$,  Eq. (\ref{KSO})] that combines the spin-orbit precession wave vector $k_{\rm so}$, the length of the weak link coupling the quantum dot to the fermionic terminals, and the frequency (squared) of the AC field. This combination offers means of varying the effect, and can also serve as a tool for measuring the spin-precession wave vector in various materials.

%%%%%%%%%%%%%%%%%%%%%%%%%%%%%%%%%%%%	
%%%%%%%%%%%%%%%%%%%%%%%%%%%%%%%%%%%%

%%%%%%%%%%%%%%%%%%%%%%%%%%%%%%%%%%%%	
%%%%%%%%%%%%%%%%%%%%%%%%%%%%%%%%%%%%

 \begin{acknowledgments}
 OEW and AA acknowledge the hospitality of the PCS at IBS, Daejeon, S. Korea, where part of this work was  supported by
IBS funding number (IBS-R024-D1). DC acknowledges the financial support to DST with project number: DST/WISE-PDF/PM-40/2023.
 \end{acknowledgments}

%%%%%%%%%%%%%%%%%%%%%%%%%%%%%%%%%%%%
%%%%%%%%%%%%%%%%%%%%%%%%%%%%%%%%%%%%%

\appendix

 %%%%%%%%%%%%%%%%%%%%%%%%%%%%%%%%%%%%	
%%%%%%%%%%%%%%%%%%%%%%%%%%%%%%%%%%%%

\section{The Keldysh Green's functions}
\label{Keldysh}

As seen in Eqs. (\ref{dNLG}), (\ref{dELG}), and (\ref{dEtun}), we need to calculate the Green's functions given in Eq. (\ref{Gkd}). Employing the Keldysh technique \cite{Jauho1994, Jauho2003, Odashima2017}), one considers the Dyson equations
\begin{align}
&J^{}_{L}e^{-i\varphi^{}_{L}(t)}G^{}_{{\bf k}d}(t,t)=\int dt' \Sigma^{}_{\bf k}(t,t')G^{}_{dd}(t',t)\ ,\nonumber\\
&J^{}_{L}G^{}_{d{\bf k}}(t,t)e^{i\varphi^{}_{L}(t)}=\int dt' G^{}_{dd}(t,t')\Sigma^{}_{\bf k}(t',t)\ ,
\label{Dy}
\end{align}
where $G_{dd}$ is the Green's function of the dot (when coupled to the terminals), and the self-energy $\Sigma^{}_{\bf k}(t,t')$ is
\begin{align}
\Sigma^{}_{\bf k}(t,t')=J^{2}_{L}e^{-i\varphi^{}_{L}(t)}g^{}_{\bf k}(t,t')e^{i\varphi^{}_{L}(t')}\ ,
\label{Sk}
\end{align}
with $g_{\bf k}$ being the Green's function of the left fermionic terminal (when decoupled from the dot).
The Dyson's equations (\ref{Dy}) pertain to all three Keldysh Green's functions, the retarded ($G^{r}_{}$), advanced ($G^{a}_{}$), and the lesser one ($G^{<}_{}$).

 %%%%%%%%%%%%%%%%%%%%%%%%%%%%%%%%%%%%	
%%%%%%%%%%%%%%%%%%%%%%%%%%%%%%%%%%%%

Dyson's equation for Green's function on the dot  reads
\begin{align}
G^{}_{dd}(t,t')&=g^{}_{d}(t,t')
+\int dt^{}_{1}\int dt^{}_{2}g^{}_{d}(t,t^{}_{1})\Sigma(t^{}_{1},t^{}_{2})G^{}_{dd}(t^{}_{2},t')\ ,
\label{Gdd}
\end{align}
where $g^{}_{d}$ is the Green's function of the decoupled dot
\begin{align}
g^{r(a)}_{d}(t,t')=\mp i\Theta (\pm t\mp t')\exp\Big [-i\varepsilon^{}_{d}(t-t'^{}_{})\Big ]\ ,
\label{grad}
\end{align}
and
\begin{align}
\Sigma=\sum_{\bf k}\Sigma^{}_{{\bf k}}+\sum_{\bf p}\Sigma^{}_{{\bf p}}\equiv \Sigma^{}_{L}+\Sigma^{}_{R}\ ,
\label{SL}
\end{align}
is the total self-energy on the dot, due to the coupling with the two electronic reservoirs.
Assuming for concreteness that the decoupled dot is empty of electrons, i.e, $\varepsilon^{}_{d}>\mu$ [$\mu=(\mu^{}_{L}+\mu^{}_{R})/2$], the corresponding lesser Green's function,  $g^{<}_{d}$, vanishes. One then finds \cite{Odashima2017}
\begin{align}
G^{<}_{dd}(t,t)=\int dt^{}_{1}\int dt^{}_{2}G^{r}_{dd}(t,t^{}_{1})\Sigma^{<}_{}(t^{}_{1},t^{}_{2})G^{a}_{dd}(t^{}_{2},t)\ .
\label{GL}
\end{align}
%%%%%%%%%%%%%%%%%%%%%%%%%%%%%%%%%%%%	
%%%%%%%%%%%%%%%%%%%%%%%%%%%%%%%%%%%%
It is rather straightforward to simplify Eq. (\ref{GL}), exploiting Langreth's rules \cite{Lan} within the wide-band approximation \cite{Odashima2017}. One finds
\begin{align}
G^{r(a)}_{dd}(t,t')=\mp i \Theta (\pm t\mp t')e^{-i\varepsilon^{}_{d}(t-t'^{}_{})\mp \Gamma (t-t')}
\ ,
\label{GDra}
\end{align}
where $\Gamma$ is the width of the Breit-Wigner resonance created on the dot due to the coupling with the electrons baths,
\begin{align}
\Gamma=\Gamma^{}_{L}+\Gamma^{}_{R}\ ,\ \ \ \Gamma^{}_{L(R)}=\pi J^{2}_{L(R)}{\cal N}^{}_{L(R)}\ ,
\end{align}
with ${\cal N}_{L(R)}$ being the (constant) density of states in the left (right) bath. The lesser Green's function on the dot is needed at equal times and takes the form \cite{OEW2017}
\begin{widetext}
\begin{align}
G^{<}_{dd}(t,t)=\int^{t}_{} dt'\int ^{t'}_{}dt''\Big (G^{r}_{dd}(t,t')[\Sigma^{<}_{L}(t',t'')+\Sigma^{<}_{R}(t',t'')]G^{a}_{dd}(t'',t)
+G^{r}_{dd}(t,t'')[\Sigma^{<}_{L}(t'',t')+\Sigma^{<}_{R}(t'',t')]G^{a}_{dd}(t',t)\Big )  \ .
\end{align}
Let us examine the part referring to the left electronic terminal (the other part is worked out similarly). Inserting  Eq. (\ref{GDra}) yields
\begin{align}
&\int^{t}_{} dt'\int ^{t'}_{}dt''[e^{-i(\varepsilon^{}_{d}-i\Gamma)(t-t')}e^{-i(\varepsilon^{}_{d}+i\Gamma)(t''-t)}
\Sigma^{<}_{L}(t',t")+e^{-i(\varepsilon^{}_{d}-i\Gamma)(t-t'')}e^{-i(\varepsilon^{}_{d}+i\Gamma)(t'-t)}
\Sigma^{<}_{L}(t'',t')]\nonumber\\
&=2i\Gamma^{}_{L}\int\frac{d\omega}{2\pi}f^{}_{L}(\omega)\int^{t}_{} dt'\int ^{t'}_{}dt''[e^{-i(\varepsilon^{}_{d}-i\Gamma)(t-t')}e^{-i(\varepsilon^{}_{d}+i\Gamma)(t''-t)}e^{-i\omega(t'-t'')-i\varphi^{}_{L}(t')+i\varphi^{}_{L}(t'')}\nonumber\\
&\hspace{2cm}+e^{-i(\varepsilon^{}_{d}-i\Gamma)(t-t'')}e^{-i(\varepsilon^{}_{d}+i\Gamma)(t'-t)}e^{-i\omega(t''-t')-i\varphi^{}_{L}(t'')+i\varphi^{}_{L}(t')}]\ ,
\end{align}
which after arranging terms becomes
\begin{align}
2i\Gamma^{}_{L}\int\frac{d\omega}{2\pi}f^{}_{L}(\omega)\int^{t}_{} dt'e^{-2\Gamma(t-t')}\int^{t'}_{}dt''[e^{-i(\omega-\varepsilon^{}_{d}-i\Gamma)(t'-t'')-i\varphi^{}_{L}(t')+i\varphi^{}_{L}(t'')}+e^{i(\omega-\varepsilon^{}_{d}+i\Gamma)(t'-t'')+i\varphi^{}_{L}(t')-i\varphi^{}_{L}(t'')}]\  .
\end{align}
It is therefore convenient to define
\begin{align}
G^{r(a)}_{L}(\omega,t)=\mp i\int ^{t}_{}dt'e^{\pm i(\omega\pm i\Gamma-\varepsilon^{}_{d})(t-t')}
e^{\pm i[\varphi^{}_{L}(t)-\varphi^{}_{L}(t')]}\ ,
\label{Graom}
\end{align}
and then
\begin{align}
&G^{<}_{dd}(t,t)=2\int ^{t}_{}dt' e^{-2\Gamma(t-t')}
 \int\frac{d\omega}{2\pi}\Big (\Gamma^{}_{L}f^{}_{L}(\omega) [G^{a}_{L}(\omega,t')-G^{r}_{L}(\omega,t')]
+\Gamma^{}_{R}f^{}_{R}(\omega)\ [G^{a}_{R}(\omega,t')-G^{r}_{R}(\omega,t')]\Big )\ .
\label{GL1}
\end{align}

%%%%%%%%%%%%%%%%%%%%%%%%%%%%%%%%%%%%	
%%%%%%%%%%%%%%%%%%%%%%%%%%%%%%%%%%%%

%%%%%%%%%%%%%%%%%%%%%%%%%%%%%%%%%%%%	
%%%%%%%%%%%%%%%%%%%%%%%%%%%%%%%%%%%%

\section{Sum rules}
\label{sumr}

%%%%%%%%%%%%%%%%%%%%%%%%%%%%%%%%%%%%
%%%%%%%%%%%%%%%%%%%%%%%%%%%%%%%%%%%%%

Here we consider the sum of all particle and energy fluxes in the junction, {\it without} resorting to the wide-band approximation.

To verify that charge is conserved in our junction, we rewrite $\dot{N}^{}_{L}(t)$, Eq. (\ref{dNLG}), in terms of the Green's function on the dot, using Eqs. (\ref{Dy}) and the definitions of the self-energy, Eqs. (\ref{Sk}) and (\ref{SL}),
\begin{align}
\dot{N}^{}_{L}(t)=&\int dt'{\rm Tr}\{\Sigma^{}_{L}(t,t')G^{}_{dd}(t',t)-G^{}_{dd}(t,t')\Sigma^{}_{L}(t',t)\}^{<}_{}
\ .
\label{IL}
\end{align}
Then the sum of all particle fluxes in the junction is
\begin{align}
\dot{N}^{}_{d}(t)&+\dot{N}^{}_{L}(t)+\dot{N}^{}_{R}(t)=-i{\rm Tr}\{dG^{<}_{dd}(t,t)/dt\}
+\int dt'{\rm Tr}\{\Sigma (t,t')G^{}_{dd}(t',t)-G^{}_{dd}(t,t')\Sigma(t',t)\}^{<}_{}\ .
\label{sumpf}
\end{align}
The time-derivative $idG^{<}_{dd}(t,t)/dt$ is obtained by using the Dyson equations
\begin{align}
&i\frac{\partial G^{a}_{dd}(t^{}_{2},t)}{\partial t}=
-\delta(t-t^{}_{2})-\varepsilon^{}_{d}G^{a}_{dd}(t^{}_{2},t)
-\int dt' G^{a}_{dd}(t^{}_{2},t')\Sigma^{a}_{}(t',t)\ ,\nonumber\\
&i\frac{\partial G^{r}_{dd}(t,t^{}_{1})}{\partial t}=
\delta(t-t^{}_{1})+\varepsilon^{}_{d}G^{r}_{dd}(t,t^{}_{1})
+\int dt' \Sigma^{r}_{}(t,t')G^{r}_{dd}(t',t^{}_{1})\ ,
\end{align}
leading to the relation
\begin{align}
i\frac{dG^{<}_{dd}(t,t)}{dt}=\int dt'{\rm Tr}\{\Sigma(t,t')G^{}_{dd}(t',t)-G^{}_{dd}(t,t')\Sigma(t',t)\}^{<}_{}\ ,
\end{align}
thus veryfing  Eq. (\ref{cc}).

%%%%%%%%%%%%%%%%%%%%%%%%%%%%%%%%%%%%
%%%%%%%%%%%%%%%%%%%%%%%%%%%%%%%%%%%%%

Consider now the sum of all energy fluxes in our junction. An alternative expression for $\dot{E}^{}_{L,{\rm tun}}(t)$ is obtained by working out the commutators in Eq. (\ref{dELtun}), yielding
\begin{align}
\dot{E}^{}_{L,{\rm tun}}(t)=&J^{}_{L}\sum_{{\bf k}}{\rm Tr}\{[\varepsilon^{}_{k}-\varepsilon^{}_{d}+\dot{\varphi}^{}_{L}(t)]
[e^{i\varphi^{}_{L}(t)}G^{<}_{d{\bf k}}(t,t)-e^{-i\varphi^{}_{L}(t)}G^{<}_{{\bf k}d}(t,t)]\}
-J^{}_{L}J^{}_{R}\sum_{{\bf k},{\bf p}}{\rm Tr}\{e^{i\varphi^{}_{L}(t)-i\varphi^{}_{R}(t)}G^{<}_{{\bf p}{\bf k}}(t,t)-e^{i(\varphi^{}_{R}(t)-i\varphi^{}_{L}(t)}G^{<}_{{\bf k}{\bf p}}(t,t)\}\ .
\label{AdEtun}
\end{align}
It follows that the sum of all energy fluxes [see Eqs.  (\ref{dEd}), (\ref{dELG}) and (\ref{dEtun})] is
\begin{align}
\dot{E}^{}_{L,{\rm tun}}(t)+\dot{E}^{}_{R,{\rm tun}}(t)+\dot{E}^{}_{d}(t)+\dot{E}^{}_{L}(t)+\dot{E}^{}_{R}(t)
&=J^{}_{L}\sum_{\bf k}{\rm Tr}\{\dot{\varphi}^{}_{L}(t)[e^{i\varphi^{}_{L}(t)}G^{<}_{d{\bf k}}(t,t)-e^{-i\varphi^{}_{L}(t)}G^{<}_{{\bf k}d}(t,t)]\}\nonumber\\
&
+J^{}_{R}\sum_{\bf p}{\rm Tr}\{\dot{\varphi}^{}_{R}(t)[e^{i\varphi^{}_{R}(t)}G^{<}_{d{\bf p}}(t,t)-e^{-i\varphi^{}_{R}(t)}G^{<}_{{\bf p}d}(t,t)]\}\ .
\label{sumE}
\end{align}
The right-hand-side in Eq. (\ref{sumE}) vanishes
when the spin-orbit interaction is static, and then energy is conserved. The time dependence of the induced  Aharonov-casher phase factors implies that energy is supplied by the AC field.

%%%%%%%%%%%%%%%%%%%%%%%%%%%%%%%%%%%%
%%%%%%%%%%%%%%%%%%%%%%%%%%%%%%%%%%%%%

\section{Explicit expressions for the fluxes}
\label{expG}

%%%%%%%%%%%%%%%%%%%%%%%%%%%%%%%%%%%%
%%%%%%%%%%%%%%%%%%%%%%%%%%%%%%%%%%%%%

\subsection{Complete time dependence}
\label{Full}

%%%%%%%%%%%%%%%%%%%%%%%%%%%%%%%%%%%%
%%%%%%%%%%%%%%%%%%%%%%%%%%%%%%%%%%%%%

The explicit expression of the particle flux, within the wide-band limit, has been derived repeatedly in the literature (see, for instance, Refs. \onlinecite{Jauho1994, Jauho2003, Odashima2017}). The new element in our study is the time dependence of the self-energy, e.g., Eq. (\ref{Sk}), embedding the Aharonov-Casher phase factors. To quantify the effect of those, we return to Eq. (\ref{IL}) and apply on it Langreth's rules,
\begin{align}
\dot{N}^{}_{L}(t)=&J^{2}_{L}\int dt'{\rm Tr}\{\Sigma^{r}_{L}(t,t')G^{<}_{dd}(t',t)+
\Sigma^{<}_{L}(t,t')G^{a}_{dd}(t',t)
-G^{r}_{dd}(t,t')\Sigma^{<}_{L}(t',t)-G^{<}_{dd}(t,t')\Sigma^{a}_{L}(t',t)\}\ .
\label{IL1}
\end{align}
In the wide-band limit, this expression takes the form
\begin{align}
\dot{N}^{}_{L}(t)=&-2i\Gamma^{}_{L}{\rm Tr}\{G^{<}_{dd}(t,t)\}+2i\Gamma^{}_{L}\int\frac{d\omega}{2\pi}f^{}_{L}(\omega){\rm Tr}\{G^{a}_{L}(\omega,t)-G^{r}_{L}(\omega,t)\}\ ,
\label{IL2}
\end{align}
where explicit expressions for the various Green's functions are to be found in Eqs.  (\ref{Graom}) and (\ref{GL1}).

%%%%%%%%%%%%%%%%%%%%%%%%%%%%%%%%%%%%
%%%%%%%%%%%%%%%%%%%%%%%%%%%%%%%%%%%%%

We next turn to the energy flux associated with the left electrons' bath, Eq. (\ref{dELG}). Exploiting the Dyson equations for
$G^{}_{{\bf k}d}$ and  $G^{}_{d{\bf k}} $ to express them in terms of $G_{dd}^{}$, and then applying the Langreth rules one finds
\begin{align}
\dot{E}^{}_{L}(t)=J^{2}_{L}\sum_{\bf k}\varepsilon^{}_{k}\int dt'{\rm Tr}&\{e^{-i\varphi^{}_{L}(t)}g^{r}_{{\bf k}}(t,t')e^{i\varphi^{}_{L}(t')}G^{<}_{dd}(t',t)+e^{-i\varphi^{}_{L}(t)}g^{<}_{{\bf k}}(t,t')e^{i\varphi^{}_{L}(t')}G^{a}_{dd}(t',t)\nonumber\\
&-G^{r}_{dd}(t,t')e^{-i\varphi^{}_{L}(t')}g^{<}_{{\bf k}}(t',t)e^{i\varphi^{}_{L}(t)}-
G^{<}_{dd}(t,t')e^{-i\varphi^{}_{L}(t')}g^{a}_{{\bf k}}(t',t)e^{i\varphi^{}_{L}(t)}\}\ .
\label{dEL1}
\end{align}
It is straightforward to work out the two terms involving $g^{<}_{\bf k}$. Indeed  \cite{Jauho1994},
\begin{align}
\sum_{\bf k}\varepsilon^{}_{k}g^{<}_{k}(t,t')=2i\frac{\Gamma^{}_{L}}{J^{2}_{L}}\int \frac{d\omega}{2\pi} e^{-i\omega (t-t')}\omega f^{}_{L}(\omega)\ ,
\end{align}
and therefore these two terms yield [see Eq. (\ref{Graom})]
\begin{align}
2i\Gamma^{}_{L}\int\frac{d\omega}{2\pi}f^{}_{L}(\omega)\omega {\rm Tr}\{G^{a}_{L}(\omega,t)-G^{r}_{L}(\omega,t)\}\ .
\label{sec}
\end{align}
The other two terms in Eq. (\ref{dEL1}) pose a difficulty, since
$\sum_{\bf k}\varepsilon^{}_{k}g^{r(a)}_{k}(t,t')$ diverges within the wide-band approximation.
However,
exploiting the relations \cite{Jauho1994}
\begin{align}
\varepsilon^{}_{k}g^{r}_{k}(t,t')=-i\frac{\partial g^{r}_{k}(t,t')}{\partial t'}-\pi\delta(t-t')\ ,\ \ \
&\varepsilon^{}_{k}g^{a}_{k}(t',t)=i\frac{\partial g^{a}_{k}(t',t)}{\partial t'}-\pi\delta(t-t')\ ,
\end{align}
shows that the terms involving the delta functions $\delta(t-t')$  are canceled,  leaving
\begin{align}
J^{2}_{L}\int dt'\sum_{\bf k}{\rm Tr}\Big \{e^{-i\varphi^{}_{L}(t)}&\Big (-i\frac{\partial g^{r}_{k}(t,t')}{\partial t'}\Big )e^{i\varphi^{}_{L}(t')}G^{<}_{dd}(t',t)
-G_{dd}^{<}(t,t')e^{-i\varphi^{}_{L}(t')}\Big (
i\frac{\partial
g^{a}_{k}(t',t)}{\partial t'}\Big )e^{i\varphi^{}_{L}(t)}\Big \}\nonumber\\
&=\Gamma^{}_{L}{\rm Tr}\Big\{\frac{dG^{<}_{dd}(t',t)}{dt'}-\frac{dG^{<}_{dd}(t,t')}{dt'}\Big \}\Big|^{}_{t'=t}
+2i\Gamma^{}_{L}{\rm Tr}\{\dot{\varphi}^{}_{L}(t)G^{<}_{dd}(t,t)\}\ .
\end{align}
Within the wide-band limit, and exploiting Eq. (\ref{GL})
\begin{align}
\Gamma^{}_{L}\Big [\frac{dG^{<}_{dd}(t',t)}{dt'}&-\frac{dG^{<}_{dd}(t,t')}{dt'}\Big]\Big|^{}_{t'=t}=2\Gamma^{}_{L}\int \frac{d\omega}{2\pi} \Big (\Gamma^{}_{L}f^{}_{L}(\omega)[G^{a}_{L}(\omega,t)+G^{r}_{L}(\omega,t)]+\Gamma^{}_{R}f^{}_{R}(\omega)[G^{a}_{R}(\omega,t)+G^{r}_{R}(\omega,t)]
\Big )-2i\Gamma^{}_{L}\varepsilon^{}_{d}G^{<}_{dd}(t,t)\ .
\end{align}
Collecting all contributions,  the time-dependent energy flux associated with the left bath is
\begin{align}
\dot{E}^{}_{L}(t)&=2i\Gamma^{}_{L}\int\frac{d\omega}{2\pi}f^{}_{L}(\omega)\omega{\rm Tr}\{G^{a}_{L}(\omega,t)-G^{r}_{L}(\omega,t)\}-2i\Gamma^{}_{L}{\rm Tr}\{G^{<}_{dd}(t,t)[\varepsilon^{}_{d}-\dot{\varphi}^{}_{L}(t)\}\nonumber\\
&+2\Gamma^{}_{L}{\rm Tr}\Big\{\Gamma^{}_{L}\int\frac{d\omega}{2\pi}f^{}_{L}(\omega) [G^{a}_{L}(\omega,t)+G^{r}_{L}(\omega,t)]+\Gamma^{}_{R}\int\frac{d\omega}{2\pi}f^{}_{R}(\omega) [G^{a}_{R}(\omega,t)+G^{r}_{R}(\omega,t)]\Big\}\ .
\label{IEL1}
\end{align}
Here again, by exploiting  Eqs. (\ref{Graom}) and (\ref{GL1}) one obtains an explicit expression for the time-dependent energy flux in the left terminal, $\dot{E}_{L}^{}(t)$.

%%%%%%%%%%%%%%%%%%%%%%%%%%%%%%%%%%%%
%%%%%%%%%%%%%%%%%%%%%%%%%%%%%%%%%%%%%

Finally, we consider the energy flux associated with the left weak link, Eq. (\ref{dEtun}), which upon using Eqs. (\ref{Dy}) becomes
\begin{align}
\dot{E}^{}_{L,{\rm tun}}(t)&=-i{\rm Tr}\{\frac{d}{dt}\int dt'[G^{}_{dd}(t,t')\Sigma^{}_{L}(t',t)+
\Sigma^{}_{L}(t,t')G^{}_{dd}(t',t)]^{<}_{}
\}\nonumber\\
&=-2i\Gamma^{}_{L}\int\frac{d\omega}{2\pi}f^{}_{L}(\omega){\rm Tr}\{\frac{d}{dt}\int^{t}_{} dt' [e^{-i(\varepsilon^{}_{d}-i\Gamma)(t-t')}e^{-i\omega(t'-t)-i\varphi^{}_{L}(t')+i\varphi^{}_{L}(t)}-
e^{-i(\varepsilon^{}_{d}+i\Gamma)(t'-t)}e^{-i\omega(t-t')-i\varphi^{}_{L}(t)+i\varphi^{}_{L}(t')}]\}\nonumber\\
&=2\Gamma^{}_{L}\int\frac{d\omega}{2\pi}f^{}_{L}(\omega){\rm Tr}\{\frac{d}{dt}[G^{a}_{}(\omega,t)+G^{r}_{}(\omega,t)]\}\ ,
\label{dEtunn}
\end{align}
where we have used the expressions in Appendix \ref{Keldysh}.

%%%%%%%%%%%%%%%%%%%%%%%%%%%%%%%%%%%%
%%%%%%%%%%%%%%%%%%%%%%%%%%%%%%%%%%%%%

\subsection{Weak Rashba interaction, small $\Omega$}
\label{int}

%%%%%%%%%%%%%%%%%%%%%%%%%%%%%%%%%%%%
%%%%%%%%%%%%%%%%%%%%%%%%%%%%%%%%%%%%%

In the present study, we confine ourselves to quantities averaged over the oscillation period of the external field, that are calculated for weak spin-orbit coupling. In that case [see Eq. (\ref{acL})],
\begin{align}
G^{r(a)}_{L}(\omega,t)=\mp i\int ^{t}_{}dt'e^{\pm i(\omega\pm i\Gamma-\varepsilon^{}_{d})(t-t')}
\Big (1-\frac{(k^{}_{\rm so}d^{}_{L})^{2}}{2}[\cos(\Omega t)-\cos(\Omega t')]^{2}\mp i\sigma^{}_{y}k^{}_{\rm so}d^{}_{L}
[\cos(\Omega t)-\cos(\Omega t')]\Big )\ .
\end{align}
Changing $t'=t-\tau$ one obtains
\begin{align}
G^{r(a)}_{L}(\omega,t)=\mp i\int_{0}^{\infty}d\tau e^{\pm i(\omega\pm i\Gamma-\varepsilon^{}_{d})\tau}
\Big (1&-\frac{(k^{}_{\rm so}d^{}_{L})^{2}}{2}\Big [\cos^{2}(\Omega t)[1-\cos(\Omega \tau)]^{2}+\sin^{2}(\Omega t)\sin^{2}(\Omega\tau)-\sin(2\Omega t)\sin(\Omega\tau)[1-\cos(\Omega\tau)]\Big ]\nonumber\\
&\mp i \sigma^{}_{y}k^{}_{\rm so}d^{}_{L}\Big [\cos(\Omega t)[1-\cos(\Omega\tau)]-\sin(\Omega t)\sin(\Omega\tau)\Big ]\Big )\ .
\label{Gra1}
\end{align}
This result implies that the period average of $\dot{E}^{}_{L,{\rm tun}}(t)$, Eq. (\ref{dEtunn}), vanishes.

%%%%%%%%%%%%%%%%%%%%%%%%%%%%%%%%%%%%
%%%%%%%%%%%%%%%%%%%%%%%%%%%%%%%%%%%%%

Period averaging Eq. (\ref{Gra1}) gives
\begin{align}
& \frac{\Omega}{2\pi}\int _{0}^{\frac{2\pi}{\Omega}}dt {\rm Tr}\{G^{r(a)}_{L}(\omega,t)\}=\mp 2 i\int_{0}^{\infty}d\tau e^{\pm i(\omega\pm i\Gamma-\varepsilon^{}_{d})\tau}\Big (1-\frac{(k^{}_{\rm so}d^{}_{L})^{2}}{2}[1-\cos(\Omega\tau)]\Big )\nonumber\\
 &\approx 2\Big (1+\frac{(k^{}_{\rm so}d^{}_{L}\Omega)^{2}}{4}\frac{\partial^{2}}{\partial\omega^{2}}\Big )\Big (\mp  i\int_{0}^{\infty}d\tau e^{\pm i(\omega\pm i\Gamma-\varepsilon^{}_{d})\tau}\Big )=2\Big (1+\frac{(k^{}_{\rm so}d^{}_{L}\Omega)^{2}}{4}\frac{\partial^{2}}{\partial\omega^{2}}\Big )\frac{1}{\omega-\varepsilon^{}_{d} \pm i\Gamma}\ ,
 \label{Tr1}
\end{align}
which yields
\begin{align}
&\frac{\Omega}{2\pi}\int _{0}^{\frac{2\pi}{\Omega}}dt {\rm Tr}\{G^{a}_{L}(\omega,t)-G^{r}_{L}(\omega,t)\}
\approx
4i\Gamma\Big (D(\omega)+\frac{(k^{}_{\rm so}d^{}_{L}\Omega)^{2}}{4}D''(\omega)\Big )\ ,\nonumber\\
&\frac{\Omega}{2\pi}\int _{0}^{\frac{2\pi}{\Omega}}dt {\rm Tr}\{G^{a}_{L}(\omega,t)+G^{r}_{L}(\omega,t)\}
\approx
4\Big ((\omega-\varepsilon^{}_{d})[D(\omega)+\frac{(k^{}_{\rm so}d^{}_{L}\Omega)^{2}}{4}D''(\omega)]+\frac{(k^{}_{\rm so}d^{}_{L}\Omega)^{2}}{2}D'(\omega)\Big )\ ,
\label{abp}
\end{align}
where [see Eq. (\ref{D})]
$D(\omega)=(\omega-\varepsilon^{}_{d})^{2}+\Gamma^{2}]^{-1}$. Likewise,
\begin{align}
\int ^{t}_{}dt' e^{-2\Gamma(t-t')}G^{r(a)}_{L}(\omega,t')=&\mp i\int^{\infty}_{0} d\tau'e^{-2\Gamma\tau'}
\int_{0}^{\infty}d\tau e^{\pm i(\omega\pm i\Gamma-\varepsilon^{}_{d})\tau}
\Big (1-\frac{(k^{}_{\rm so}d^{}_{L})^{2}}{2}
\Big [[\cos(\Omega t)\cos(\Omega\tau')+\sin(\Omega t)\sin(\Omega\tau')]^{2}\nonumber\\
&\times[1-\cos(\Omega \tau)]^{2}+[\sin(\Omega t)\cos(\Omega\tau')-\cos(\Omega t)\sin(\Omega\tau')]^{2}\sin^{2}(\Omega\tau)\nonumber\\
&
-[\sin(2\Omega t)\cos(2\Omega\tau')-\cos(2\Omega t)\sin(2\Omega\tau')]\sin(\Omega\tau)[1-\cos(\Omega\tau)]\Big ]\nonumber\\
&\hspace{-3.5cm}\mp i \sigma^{}_{y}k^{}_{\rm so}d^{}_{L}\Big [[\cos(\Omega t)\cos(\Omega\tau')+\sin(\Omega t)\sin(\Omega\tau')][1-\cos(\Omega\tau)]-[\sin(\Omega t)\cos(\Omega\tau')-\cos(\Omega t)\sin(\Omega\tau')]\sin(\Omega\tau)\Big ]\Big )\ ,
\end{align}
leading to
\begin{align}
\frac{\Omega}{2\pi}\int ^{\frac{2\pi}{\Omega}}_{0}dt\int^{t}dt'e^{-2\Gamma(t-t')}{\rm Tr}
\{G^{a}_{L}(\omega.t')-G^{r}_{L}(\omega.t')\}\approx
2i\Big (D(\omega)+\frac{(k^{}_{\rm so}d^{}_{L}\Omega)^{2}}{4}D''(\omega)\Big )\ ,
\label{TrGam}
\end{align}
and
\begin{align}
&-\frac{\Omega}{2\pi}\int ^{\frac{2\pi}{\Omega}}_{0}dt\int^{t}dt'e^{-2\Gamma(t-t')}{\rm Tr}
\{[G^{a}_{L}(\omega.t')-G^{r}_{L}(\omega.t')]\sigma^{}_{y}\Omega\sin(\Omega t)\}\nonumber\\
&=-k^{}_{\rm so}d^{}_{L}\Omega\int _{0}^{\infty}d\tau' e^{-2\Gamma\tau'}\int_{0}^{\infty }d\tau
\Big (e^{-i(\omega-i\Gamma-\varepsilon^{}_{d})\tau}-e^{i(\omega+i\Gamma-\varepsilon^{}_{d})\tau}\Big )\cos(\Omega\tau')\sin(\Omega\tau)
\approx -\frac{\Omega^{2}}{2\Gamma}k^{}_{\rm so}d^{}_{L}\int_{0}^{\infty }d\tau\tau
\Big [e^{-i(\omega-i\Gamma-\varepsilon^{}_{d})\tau}-e^{i(\omega+i\Gamma-\varepsilon^{}_{d})\tau}
\Big ]\Big )\nonumber\\
&\hspace{3cm}=-i\frac{\Omega^{2}}{2\Gamma}k^{}_{\rm so}d^{}_{L}\frac{d}{d\omega}
\int_{0}^{\infty }d\tau
\Big (e^{-i(\omega-i\Gamma-\varepsilon^{}_{d})\tau}+e^{i(\omega+i\Gamma-\varepsilon^{}_{d})\tau}\Big )%\nonumber\\
=-i\Omega^{2}k^{}_{\rm so}d^{}_{L}D'(\omega)\ .
\label{GLsig}
\end{align}
Exploiting these results leads to Eq. (\ref{ILav}) for the averaged particle flux.

%%%%%%%%%%%%%%%%%%%%%%%%%%%%%%%%%%%%
%%%%%%%%%%%%%%%%%%%%%%%%%%%%%%%%%%%%%

The period average of $\dot{E}_{L}(t)$, Eq. (\ref{IEL1}), is found as follows.  Employing the second of Eqs. (\ref{abp}),
\begin{align}
&2\Gamma^{}_{L}\frac{\Omega}{2\pi}\int _{0}^{\frac{2\pi}{\Omega}}dt
{\rm Tr}\Big\{\Gamma^{}_{L}\int\frac{d\omega}{2\pi}f^{}_{L}(\omega) [G^{a}_{L}(\omega,t)+G^{r}_{L}(\omega,t)]+\Gamma^{}_{R}\int\frac{d\omega}{2\pi}f^{}_{R}(\omega) [G^{a}_{R}(\omega,t)+G^{r}_{R}(\omega,t)]\Big\}\nonumber\\
&=8\Gamma^{}_{L}\int\frac{d\omega}{2\pi}\Big ([\Gamma^{}_{L}f^{}_{L}(\omega)+\Gamma^{}_{R}f^{}_{R}(\omega)](\omega-\varepsilon^{}_{d})D(\omega)
+\frac{(k^{}_{\rm so}\Omega)^{2}}{4}[\Gamma^{}_{L}f^{}_{L}(\omega)d^{2}_{L}+\Gamma^{}_{R}f^{}_{R}(\omega)d^{2}_{R}]\frac{d^{2}}{d\omega^{2}}[(\omega-\varepsilon^{}_{d})D(\omega)]\Big)\ ,
\end{align}
while  Eq.  (\ref{TrGam})  gives
\begin{align}
-2i\Gamma^{}_{L}\varepsilon^{}_{d}{\rm Tr}\{G^{<}_{dd}(t,t)\}=8\Gamma^{}_{L}\varepsilon^{}_{d}\int\frac{d\omega}{2\pi}\Big ([\Gamma^{}_{L}f^{}_{L}(\omega)+\Gamma^{}_{R}f^{}_{R}(\omega)]D(\omega)+\frac{(k^{}_{\rm so}\Omega)^{2}}{4}
[\Gamma^{}_{L}f^{}_{L}(\omega)d^{2}_{L}+\Gamma^{}_{R}f^{}_{R}(\omega)d^{2}_{R}]
D''(\omega)\Big)\ ,
\end{align}
so that the sum of these two contributions is just
\begin{align}
8\Gamma^{}_{L}\int\frac{d\omega}{2\pi}\Big ([\Gamma^{}_{L}f^{}_{L}(\omega)+\Gamma^{}_{R}f^{}_{R}(\omega)]\omega D(\omega)+\frac{(k^{}_{\rm so}\Omega)^{2}}{4}[\Gamma^{}_{L}f^{}_{L}(\omega)d^{2}_{L}+\Gamma^{}_{R}f^{}_{R}(\omega)d^{2}_{R}][\omega D''(\omega)+2D'(\omega)]\Big )\ .
\end{align}
Adding to it the contribution of the first term on the right-hand side of Eq. (\ref{IEL1}),
\begin{align}
2i\Gamma^{}_{L}\int\frac{d\omega}{2\pi}f^{}_{L}(\omega)\omega\frac{\Omega}{2\pi}\int _{0}^{\frac{2\pi}{\Omega}}dt
{\rm Tr}\{G^{a}_{L}(\omega,t)-G^{r}_{L}(\omega,t)\}=2(2i\Gamma^{}_{L})(2i\Gamma)\int\frac{d\omega}{2\pi}f^{}_{L}(\omega)\omega\Big (D(\omega)+\frac{(k^{}_{\rm so}\Omega)^{2}}{4}d^{2}_{L}D''(\omega)\Big)\ ,
\end{align}
one obtains
\begin{align}
&8\Gamma^{}_{L}\Gamma^{}_{R}\int\frac{d\omega}{2\pi}\Big ([f^{}_{R}(\omega)-f^{}_{L}(\omega)]\omega D(\omega)\nonumber\\
&+\frac{(k^{}_{\rm so}\Omega)^{2}}{4}[f^{}_{R}(\omega)d^{2}_{R}-f^{}_{L}(\omega)d^{2}_{L}]\omega D''(\omega)\Big )+4\Gamma^{}_{L}(k^{}_{\rm so}\Omega)^{2}\int\frac{d\omega}{2\pi}D'(\omega)[\Gamma^{}_{L}f^{}_{L}(\omega)d^{2}_{L}+\Gamma^{}_{R}f^{}_{R}(\omega)d^{2}_{R}]\ .
\label{IELa}
\end{align}
It remains to consider
\begin{align}
-2i\Gamma^{}_{L}\frac{\Omega}{2\pi}\int _{0}^{\frac{2\pi}{\Omega}}dt
{\rm Tr}\{\dot{\varphi}^{}_{L}(t)G^{<}_{dd}(t,t)\}=-2i\Gamma^{}_{L}k^{}_{\rm so}d^{}_{L}\frac{\Omega}{2\pi}\int _{0}^{\frac{2\pi}{\Omega}}dt
{\rm Tr}\{\sigma^{}_{y}\Omega\sin(\Omega t)G^{<}_{dd}(t,t)\}\ ,
\end{align}
which using  Eqs. (\ref{GL1}) and (\ref{GLsig}) becomes
\begin{align}
-2\Gamma^{}_{L}(\Omega k^{}_{\rm so})^{2}d^{}_{L}\int\frac{d\omega}{2\pi}D'(\omega)[2\Gamma^{}_{L}f^{}_{L}(\omega)d^{}_{L}
-2\Gamma^{}_{R}f^{}_{R}(\omega)d^{}_{R}]\ .
\label{IELb}
\end{align}

%%%%%%%%%%%%%%%%%%%%%%%%%%%%%%%%%%%%
%%%%%%%%%%%%%%%%%%%%%%%%%%%%%%%%%%%%%

\section{Linear-response coefficients}

\label{LR}

%%%%%%%%%%%%%%%%%%%%%%%%%%%%%%%%%%%%
%%%%%%%%%%%%%%%%%%%%%%%%%%%%%%%%%%%%%

The coefficients that determine the thermoelectric transport in the linear-response regime are given by
Eq. (\ref{In}), reproduced here for convenience
\begin{align}
I^{}_{n}=I^{(0)}_{n}\Big (1+K^{}_{\rm so}I^{(2)}_{n}/I^{(0)}_{n}\Big )\ ,\ \ n=0,1,2\ ,
\label{D1}
\end{align}
where $I^{(\ell)}_{n}$, Eq. (\ref{sch}), expressed in dimensionless units, is
\begin{align}
I^{(\ell)}_{n}[\beta(\varepsilon^{}_{d}-\mu)]=\beta^{\ell-n}_{}\frac{4}{\pi}\beta^{2}\Gamma^{}_{L}\Gamma^{}_{R}
\int_{-\infty}^{\infty}dxF^{}_{n}(x)%\frac{x^{n}_{}}{4\cosh^{2}(x/2)}
\frac{d^{\ell}_{}}{dx^{\ell}_{}}\frac{1}{[\beta(\varepsilon^{}_{d}-\mu)-x]^{2}+(\beta\Gamma)^{2}}\ ,\ \ F^{}_{n}(x)=
\frac{x^{n}_{}}{4\cosh^{2}(x/2)}\ ,\ \
\ell=0,2\ .
\label{D2}
\end{align}

For plotting Figs. \ref{Sco} and \ref{Kap}
it is convenient to use dimensionless parameters. The prefactor in Eq (\ref{D2}) implies that $K^{}_{\rm so}$ is multiplied by $\beta^{2}$ times a dimensionless ratio of integrals. However, multiplying this ratio by $[\beta(\varepsilon^{}_{d}-\mu)]^{2}$ yields an expansion in $K^{}_{\rm so}/(\varepsilon^{}_{d}-\mu)^{2}$, which is temperature-independent. Fixing this parameter at $1/16$ (see end of Sec. \ref{Intro}) implies that the temperature of the results is included in the abscissa $\beta(\varepsilon^{}_{d}-\mu)$ and in $\beta\Gamma$ (fixed at the small value $0.2$). Since we include only a single level on the dot, we must have $\Gamma<<\varepsilon^{}_{d}-\mu$ and $\beta(\varepsilon^{}_{d}-\mu)>>1$. Therefore, one should ignore the small values of $\beta(\varepsilon^{}_{d}-\mu)$ in the figures.

%%%%%%%%%%%%%%%%%%%%%%%%%%%%%%%%%%%%
%%%%%%%%%%%%%%%%%%%%%%%%%%%%%%%%%%%%%

\subsection{High temperature approximation}

\label{ht}

Focusing on thermoelectric transport in quantum dots (in the absence of AC driving and spin-orbit coupling, i.e., for $K^{}_{\rm so}=0$), a ubiquitous approximation in the literature \cite{Beenakker1992, Edwards1995} (see also Ref. \onlinecite{Kouwenhoven1997}), is based on a (relative) high-temperature approximation, assuming that $\beta\Gamma<<1$,
\begin{align}
\frac{1}{[\beta(\varepsilon^{}_{d}-\mu)-x]^{2}+(\beta\Gamma)^{2}}\approx\frac{\pi}{\beta\Gamma}\delta[\beta(\varepsilon^{}_{d}-\mu)-x]\ .
\end{align}
When applied to the integrals resulting from the effect of the spin-orbit coupling, it yields
\begin{align}
I^{(\ell)}_{n}[\beta(\varepsilon^{}_{d}-\mu)]=\beta^{\ell-n}_{}\frac{4\beta^{2}\Gamma^{}_{L}\Gamma^{}_{R}}{\beta\Gamma}
\frac{d^{\ell}_{}}{du^{\ell}_{}}F^{}_{n}(u)%\frac{x^{n}_{}}{4\cosh^{2}(x/2)}
\Big|^{}_{u=\beta(\varepsilon^{}_{d}-\mu)}\ ,\ \ \ell=0,2\ .
\end{align}
%%%%%%%%%%%%%%%%%%%%%%%%%%%%%%%%%%%%
%%%%%%%%%%%%%%%%%%%%%%%%%%%%%%%%%%%%%
Using these results in Eq. (\ref{D1}), one finds
\begin{align}
I^{}_{0}&\approx\frac{4\beta\Gamma^{}_{L}\Gamma^{}_{R}}{\Gamma}
F^{}_{n=0}(u)\Big (1+\beta^{2}K^{}_{\rm so}F''^{}_{n=0}(u)/F^{}_{n=0}(u)\Big )\Big|^{}_{u=\beta(\varepsilon^{}_{d}-\mu)}\ ,\nonumber\\
I^{}_{1}&\approx\frac{4\Gamma^{}_{L}\Gamma^{}_{R}}{\Gamma}
F^{}_{n=1}(u)\Big (1+\beta^{2}K^{}_{\rm so}F''^{}_{n=1}(u)/F^{}_{n=1}(u)\Big )\Big|^{}_{u=\beta(\varepsilon^{}_{d}-\mu)}\ ,\nonumber\\
I^{}_{2}&\approx\frac{4\Gamma^{}_{L}\Gamma^{}_{R}}{\beta\Gamma}
F^{}_{n=2}(u)\Big (1+\beta^{2}K^{}_{\rm so}F''^{}_{n=2}(u)/F^{}_{n=2}(u)\Big )\Big|^{}_{u=\beta(\varepsilon^{}_{d}-\mu)}\ .
\end{align}

%%%%%%%%%%%%%%%%%%%%%%%%%%%%%%%%%%%%
%%%%%%%%%%%%%%%%%%%%%%%%%%%%%%%%%%%%%

The Seebeck coefficient is determined by
\begin{align}
S=\frac{I^{}_{1}}{I^{}_{0}}=(\varepsilon^{}_{d}-\mu)\frac{1+\beta^{2}K^{}_{\rm so}F''^{}_{n=1}(u)/F^{}_{n=1}(u)}{1+\beta^{2}K^{}_{\rm so}F''^{}_{n=0}(u)/F^{}_{n=0}(u)}\Big|^{}_{u=\beta(\varepsilon^{}_{d}-\mu)}\ ,
\end{align}
with
\begin{align}
F''^{}_{n=1}(u)/F^{}_{n=1}(u)-F''^{}_{n=0}(u)/F^{}_{n=0}(u)=2F'(u)/[uF(u)]\ ,
\end{align}
which is negative. The effect of the spin-orbit coupling is to reduce the Seebeck coefficient.

%%%%%%%%%%%%%%%%%%%%%%%%%%%%%%%%%%%%
%%%%%%%%%%%%%%%%%%%%%%%%%%%%%%%%%%%%%

The thermal conductance is given by
\begin{align}
\kappa^{}_{\rm e}=I^{}_{2}-\frac{I^{2}_{1}}{I^{}_{0}}=\frac{4\Gamma^{}_{L}\Gamma^{}_{R}}{\beta\Gamma}\Big\{
F^{}_{n=2}(u)\Big (1+\beta^{2}K^{}_{\rm so}\frac{F''^{}_{n=2}(u)}{F^{}_{n=2}(u)}\Big )-\frac{F^{2}_{n=1}(u)}{F^{}_{n=0}(u)}
\frac{1+2\beta^{2}K^{}_{\rm so}F''^{}_{n=1}(u)/F^{}_{n=1}(u)}{1+\beta^{2}K^{}_{\rm so}F''^{}_{n=0}(u)/F^{}_{n=0}(u)}\Big\}
\Big|^{}_{u=\beta(\varepsilon^{}_{d}-\mu)}\ ,
\end{align}
with
\begin{align}
F^{}_{n=2}(u)&=F^{2}_{n=1}(u)/F^{}_{n=0}(u)\ ,\nonumber\\
\frac{F''^{}_{n=2}(u)}{F^{}_{n=2}(u)}&-2\frac{F''^{}_{n=1}(u)}{F^{}_{n=1}(u)}
+\frac{F''^{}_{n=0}(u)}{F^{}_{n=0}(u)}=2/u^{2}\ .
\end{align}
Hence,
\begin{align}
\kappa^{}_{\rm e}=\frac{8\Gamma^{}_{L}\Gamma^{}_{R}}{\beta\Gamma}\frac{K^{}_{\rm so}}{(\varepsilon^{}_{d}-\mu)^{2}}\ .
\end{align}
It follows that in this approximation, the effect of the spin-orbit coupling is detrimental-- it reduces the Seebeck coefficient and enhances the thermal electronic conductance.

%%%%%%%%%%%%%%%%%%%%%%%%%%%%%%%%%%%%
%%%%%%%%%%%%%%%%%%%%%%%%%%%%%%%%%%%%%

Unfortunately, we had difficulties with the numerical integrations (using Mathematica) for $\beta\Gamma<0.2$. However, we note that all our numerical integrals do approach their high-temperature approximations as $\beta\Gamma$ decreases. Values of $\beta\Gamma<0.2$ may shift the high peaks in the plots and modify our quantitative conclusions.

%%%%%%%%%%%%%%%%%%%%%%%%%%%%%%%%%%%%
%%%%%%%%%%%%%%%%%%%%%%%%%%%%%%%%%%%%%

\subsection{Low temperature approximation}

\label{Lht}

%%%%%%%%%%%%%%%%%%%%%%%%%%%%%%%%%%%%
%%%%%%%%%%%%%%%%%%%%%%%%%%%%%%%%%%%%%

On the other hand, one may assume \cite{OEW2020} low enough temperatures, such that $\beta(\varepsilon^{}_{d}-\mu)>1$. Expanding the second factor in the integrals (\ref{D2}) to linear order in $x$, one obtains
\begin{align}
\frac{1}{[\beta(\varepsilon^{}_{d}-\mu)-x]^{2}+(\beta\Gamma)^{2}}&\approx
\frac{1}{[\beta(\varepsilon^{}_{d}-\mu)]^{2}+(\beta\Gamma)^{2}}+\frac{2x\beta(\varepsilon^{}_{d}-\mu)}{\{[\beta(\varepsilon^{}_{d}-\mu)]^{2}+(\beta\Gamma)^{2}\}^{2}_{}}%\equiv [D(x)-xD'(x)]\Big |^{}_{x=\beta(\varepsilon^{}_{d}-\mu)}
\ ,\ \ \ {\rm for}\ \ \ell=0\ ,
\end{align}
and
\begin{align}
\frac{d^{2}}{dx^{2}}\frac{1}{[\beta(\varepsilon^{}_{d}-\mu)-x]^{2}+(\beta\Gamma)^{2}}&=
\frac{6[\beta(\varepsilon^{}_{d}-\mu)-x]^{2}-2(\beta\Gamma)^{2}}
{\{[\beta(\varepsilon^{}_{d}-\mu)-x]^{2}+(\beta\Gamma)^{2}\}^{3}_{}}
\approx\frac{6[\beta(\varepsilon^{}_{d}-\mu)]^{2}-2(\beta\Gamma)^{2}}
{\{[\beta(\varepsilon^{}_{d}-\mu)]^{2}+(\beta\Gamma)^{2}\}^{3}_{}}\nonumber\\
&+24\beta(\varepsilon^{}_{d}-\mu)x\frac{[\beta(\varepsilon^{}_{d}-\mu)]^{2}-(\beta\Gamma)^{2}}
{\{[\beta(\varepsilon^{}_{d}-\mu)]^{2}+(\beta\Gamma)^{2}\}^{4}_{}}
%&\equiv [D''(x)-xD'''(x)]\Big |^{}_{x=\beta(\varepsilon^{}_{d}-\mu)}
\ ,\ \ \ {\rm for}\ \ \ell=2\ .
\end{align}
Then, using
\begin{align}
\int_{-\infty}^{\infty}dx\frac{1}{4\cosh^{2}(x/2)}=1\ ,\ \ \ \int_{-\infty}^{\infty}dx\frac{x^{2}_{}}{4\cosh^{2}(x/2)}=\pi^{2}/3\ ,
\end{align}
one finds
\begin{align}
I^{}_{0}&=\frac{4}{\pi}\beta^{2}\Gamma^{}_{L}\Gamma^{}_{R}D(u)\Big (1+\beta^{2}K^{}_{\rm so}D''(u)/D(u)\Big )\Big |^{}_{u=\beta(\varepsilon^{}_{d}-\mu)}\ ,\nonumber\\
I^{}_{1}&=-\frac{4\pi}{3}\beta\Gamma^{}_{L}\Gamma^{}_{R}D'(u)\Big (1+\beta^{2}K^{}_{\rm so}D'''(u)/D'(u)\Big )\Big |^{}_{u=\beta(\varepsilon^{}_{d}-\mu)}\ ,\nonumber\\
I^{}_{2}&=\frac{4\pi}{3}\Gamma^{}_{L}\Gamma^{}_{R}D(u)\Big (1+\beta^{2}K^{}_{\rm so}D''(u)/D(u)\Big )\Big |^{}_{u=\beta(\varepsilon^{}_{d}-\mu)}\ .
\end{align}
[Note that the derivatives or $D$ are changed from being with respect to $x$ to being with respect to $\beta(\varepsilon^{}_{d}-\mu)$.]

%%%%%%%%%%%%%%%%%%%%%%%%%%%%%%%%%%%%
%%%%%%%%%%%%%%%%%%%%%%%%%%%%%%%%%%%%%

%%%%%%%%%%%%%%%%%%%%%%%%%%%%%%%%%%%%	
%%%%%%%%%%%%%%%%%%%%%%%%%%%%%%%%%%%%

The Seebeck coefficient is
\begin{align}
S=-\frac{\pi^{2}}{3\beta}\frac{D'(u)}{D(u)}\Big (1+\beta^{2}K^{}_{\rm so}\Big [\frac{D'''(u)}{D'(u)}-\frac{D''(u)}{D(u)}\Big]\Big )\Big |^{}_{u=\beta(\varepsilon^{}_{d}-\mu)}=\frac{2\pi^{2}}{3\beta}\frac{u}{u^{2}+(\beta\Gamma)^{2}}\Big (1+\beta^{2}K^{}_{\rm so}\frac{6u^{2}-10(\beta\Gamma)^{2}}{[u^{2}+(\beta\Gamma)^{2}]^{2}}\Big )\Big |^{}_{u=\beta(\varepsilon^{}_{d}-\mu)}\ .
\end{align}
In this approximation, the effect of the spin-orbit coupling is to enhance the Seebeck coefficient.

%%%%%%%%%%%%%%%%%%%%%%%%%%%%%%%%%%%%
%%%%%%%%%%%%%%%%%%%%%%%%%%%%%%%%%%%%%

The electronic thermal conductance comprises the combination
\begin{align}
\kappa^{}_{\rm e}&=I^{}_{2}-I^{2}_{1}/I^{}_{0} .
\end{align}
This expression obviously should be positive. As our approximation is based on $\beta(\varepsilon^{}_{d}-\mu)>1$, this implies that for $K^{}_{\rm so}=0$,
\begin{align}
\Big (D^{2}_{}(u)-\frac{\pi^{2}}{3}[D'(u)]^{2}\Big )\Big |^{}_{u=\beta(\varepsilon^{}_{d}-\mu)}>0\ \ \Rightarrow\ \
\frac{4\pi^{2}}{3}\frac{u^{2}}{[u^{2}+(\beta\Gamma)^{2}]^{2}}\Big |^{}_{u=\beta(\varepsilon^{}_{d}-\mu)}<1\ .
\label{posk}
\end{align}
In the presence of the spin-orbit coupling, at linear order in $\beta^{2}K^{}_{\rm so}$, the thermal conductance is
\begin{align}
\kappa^{}_{\rm e}
&=\frac{4\pi}{3}\Gamma^{}_{L}\Gamma^{}_{R}\Big\{D(u)\Big (1+\beta^{2}K^{}_{\rm so}\frac{D''(u)}{D(u)}\Big)-\frac{\pi^{2}}{3}
\frac{[D'(u)]^{2}}{D^{}(u)}\Big (1+\beta^{2}K^{}_{\rm so}\Big [2\frac{D'''(u)}{D'(u)}-\frac{D''(u)}{D(u)}\Big]\Big )\Big\}\Big |^{}_{u=\beta(\varepsilon^{}_{d}-\mu)}\nonumber\\
&=\frac{4\pi}{3}\Gamma^{}_{L}\Gamma^{}_{R}\Big \{\frac{1}{u^{2}+(\beta\Gamma)^{2}}\Big (1+\beta^{2}K^{}_{\rm so}\frac{6u^{2}-2(\beta\Gamma)^{2}}{[u^{2}+(\beta\Gamma)^{2}]^{2}}\Big )-\frac{\pi^{2}}{3}\frac{4u^{2}}{[u^{2}+(\beta\Gamma)^{2}]^{3}}\Big (1+\beta^{2}K^{}_{\rm so}\frac{18u^{2}-22(\beta\Gamma)^{2}}{[u^{2}+(\beta\Gamma)^{2}]^{2}}\Big )\Big\}\Big |^{}_{u=\beta(\varepsilon^{}_{d}-\mu)}\ .
\end{align}
It follows that the effect of the spin-orbit coupling on the electronic thermal conductance depends on parameters. For example, ignoring $(\beta\Gamma)$ compared to $u$ one finds that $\kappa^{}_{\rm e}$ will be reduced by the spin-orbit coupling when $1<4\pi/u^{2}< 3$. For larger values of $u$ the spin-orbit coupling enhances $\kappa^{}_{\rm e}$ and then the figure of merit is deteriorated.

%%%%%%%%%%%%%%%%%%%%%%%%%%%%%%%%%%%%
%%%%%%%%%%%%%%%%%%%%%%%%%%%%%%%%%%%%%

%%%%%%%%%%%%%%%%%%%%%%%%%%%%%%%%%%%%	
%%%%%%%%%%%%%%%%%%%%%%%%%%%%%%%%%%%%

%%%%%%%%%%%%%%%%%%%%%%%%%%%%%%%%%%%%
%%%%%%%%%%%%%%%%%%%%%%%%%%%%%%%%%%%%%

\end{widetext}
%%%%%%%%%%%%%%%%%%%%%%%%%%%%%%%%%%%%	
%%%%%%%%%%%%%%%%%%%%%%%%%%%%%%%%%%%%

%%%%%%%%%%%%%%%%%%%%%%%%%%%%%%%%%%%%	
%%%%%%%%%%%%%%%%%%%%%%%%%%%%%%%%%%%%


\begin{thebibliography}{99}
\bibitem{comsoi} This simple form of the AC phase factor is strictly justified only up to quadratic order in $k^{}_{\rm so}d$. Therefore our results are limited to $k^{}_{\rm so}d \ll 1$.%%%%%%%%%%%%%%%%%%%%%%%%%%%%%%%%%%%%%
\bibitem{comcur}
The detailed derivation omits the electronic charge $e$ from the definition of the charge current, which is then referred to as `particle current' (or flux).
\end{thebibliography}
\end{document}